\documentclass[12pt]{article}
\usepackage{amsmath}
\usepackage{amssymb}
\usepackage{graphicx}
\usepackage{setspace}
\usepackage{graphicx}
\usepackage{epsfig}
\usepackage{amsfonts}
\usepackage{dsfont}
\usepackage{bm}
\usepackage{color}
\usepackage{comment}
\usepackage[dvipsnames]{xcolor}
\usepackage[authoryear,round]{natbib}
\usepackage{url}
\usepackage{pgfplots}
\usepackage[hidelinks]{hyperref}
\hypersetup{
  colorlinks   = true, 
  urlcolor     = blue, 
  linkcolor    = blue, 
  citecolor   = blue 
}
\usepackage[T1]{fontenc}
\usepackage[cal=boondoxo]{mathalfa}

\usepackage{soul}

\newtheorem{corollary}{\bf Corollary}
\newtheorem{definition}{\bf Definition}

\newtheorem{proposition}{\bf Proposition}
\newtheorem{theorem}{\bf Theorem}
\newtheorem{lemma}{\bf Lemma}

\newenvironment{proof}[1][Proof]{\noindent\textbf{#1} }{\ \rule{0.5em}{0.5em}}

\renewcommand{\ge}{\geqslant}
\renewcommand{\le}{\leqslant}

\pgfplotsset{compat=1.18}
\DeclareMathOperator{\dd}{\textrm{d}\!}
\allowdisplaybreaks

\begin{document}
\title{Robust prizes in all-pay contests with loss aversion}
\author{Dmitry Ryvkin\thanks{Economics Discipline Group, School of Economics, Finance and Marketing, RMIT University, 445 Swanston Street, Melbourne 3000, VIC, Australia, E-mail: \url{d.ryvkin@gmail.com}, ORCID: 0000-0001-9314-5441.} \and Qin Wu\thanks{Economics Discipline Group, School of Economics, Finance and Marketing, RMIT University, 445 Swanston Street, Melbourne 3000, VIC, Australia, E-mail: \url{qin.wu@rmit.edu.au}, ORCID: 0000-0002-1024-5058.}}
\date{This version: \today}

\maketitle

\begin{abstract}
\noindent We study robust prize design in all-pay contests with loss-averse agents when the principal does not know the agents' convex cost of effort, subject to an aggregate productivity constraint. The robustly optimal prize structure awards equal prizes to all contestants except the one ranked last. The optimal bottom prize is zero under moderate loss aversion, but becomes positive and increases with loss aversion when loss aversion is sufficiently strong. Under strong loss aversion, the robust solution involves partial participation in the contest.

\bigskip


\end{abstract}

\newpage
\onehalfspacing

\section{Introduction}
\label{sec:introduction}

Contests are widely used to motivate effort within organizations. Examples include sales competitions and promotion tournaments in firms, and teaching and research awards in universities. Participation in these contests is often voluntary. An academic may, for example, decide at the start of the year to pursue a teaching award and devote additional effort to improving his teaching throughout the year. Alternatively, he may choose not to pursue the award, anticipating the disappointment of an unsuccessful attempt. This behavior can arise under expectation-based loss aversion, when the psychological losses from unfavorable outcomes outweigh the gains from favorable ones. How should a manager allocate rewards when they affect both the willingness to participate and the effort of those who decide to enter competition, in the presence of loss aversion?

To study this question, we consider an all-pay contest with loss-averse agents. Additionally, we assume the manager faces \emph{ambiguity} about the agents' effort costs. While managers may know something about how employees respond to incentives, the assumption that they know the entire effort cost function is rather demanding. We assume instead that the manager knows that costs are increasing and convex and satisfy a lower bound on overall average productivity, but otherwise the cost function is unrestricted. We adopt a \emph{robust design} approach: The manager chooses prizes to maximize expected effort under the least favorable admissible cost function. For each cost function, a prize schedule is evaluated by the largest expected effort attainable in a symmetric equilibrium.

We show that a ``punish-the-bottom'' prize schedule is robustly optimal: All agents except the one ranked last receive a common prize, while the last-ranked agent receives a lower prize. Under moderate loss aversion, the bottom prize is zero and the budget is divided equally among the remaining ranks. Under strong loss aversion, the optimal bottom prize becomes strictly positive and increases with loss aversion, while the common top prize falls. Thus, it is always optimal to punish the agent finishing last, but loss aversion changes the optimal size of the penalty. The optimal bottom prize, when it is positive, does not induce full participation. Under strong loss aversion, the robust effort guarantee is attained in a partial-entry equilibrium, and the entry probability in this equilibrium decreases with loss aversion, even though the bottom prize is increasing.

To see why a positive bottom prize can be useful, consider a prize schedule with a bottom prize $b\ge 0$ and a common top prize $b+d$, for some $d>0$. An agent who receives the top prize with probability $P$ obtains an expected benefit of $b+[P-\theta P(1-P)]d$, where $\theta>0$ measures the strength of loss aversion. The term $Pd$ is the expected monetary gain over the bottom prize, while $\theta P(1-P)d$ is the net psychological loss from comparisons between the two possible outcomes. When $\theta>1$, the expression in brackets is negative for sufficiently small $P>0$. In this case, the agent would prefer to receive $b$ with certainty rather than face the prize lottery. Raising $b$ makes participation more attractive, but, under a fixed budget, it also reduces $d$. Therefore, the manager must weigh the participation benefit and the reduction in psychological losses against the weaker effort incentives from a smaller prize differential.

Our analysis separates equilibrium behavior from ambiguity about effort costs. Since costs are strictly increasing, ranking agents by effort is equivalent to ranking them by cost \emph{expenditure}. Equilibrium in expenditure space is, therefore, independent of the cost function. We first maximize expected expenditure, obtaining a bound that applies to every symmetric equilibrium under every prize schedule such that an equilibrium exists. Because a linear cost function belongs to the ambiguity set, this expenditure bound also provides an upper bound on the effort that any prize schedule can guarantee. We then construct an equilibrium under the proposed optimal prize schedule that attains the expenditure bound and show that the linear cost function is least favorable for this equilibrium. The resulting effort guarantee attains the upper bound, which proves robust optimality.

For managers, our results provide a simple prescription that requires limited knowledge of effort costs: Award a common prize to all but the last-ranked participant, and soften the penalty for finishing last when loss aversion is strong. There is a trade-off between encouraging participation and maintaining effort incentives. Under a fixed budget, a higher bottom prize makes entry more attractive but leaves fewer resources for rewarding better performance. A manager seeking to maximize effort should, therefore, consider both how many agents participate and how strongly they are motivated once they enter. Under strong loss aversion, the robust effort guarantee is attained with a positive bottom prize \emph{and} partial participation. Thus, maximum participation in competition for awards, for example, should not always be a design goal; and a reduction in entry does not imply that rewards are not optimally structured.

\paragraph{Related literature}
Our paper contributes to the literature on the optimal allocation of prizes in contests and tournaments. For agents with standard preferences, this question has been studied, for example, by \cite{Moldovanu-Sela:2001}, \cite{Fang-et-al:2020}, and \cite{Drugov-Ryvkin:2020_prizes_JET} for all-pay contests under incomplete information, all-pay contests under complete information, and rank-order tournaments with stochastic performance, respectively. The closest benchmark for our analysis is \cite{Fang-et-al:2020}, who show that, with identical agents and convex effort costs, awarding equal prizes to all agents except the one ranked last maximizes expected effort. We show that this prize schedule remains optimal for any convex cost function under moderate expectation-based loss aversion. Under strong loss aversion, however, participation becomes a substantive consideration, and the robustly optimal schedule awards a strictly positive bottom prize. Thus, the punish-the-bottom structure is preserved, but the severity of the punishment decreases with loss aversion.

Several papers considered contests and tournaments with loss-averse agents using different specifications of reference points. \cite{Gill-Stone:2010} explore two-player tournaments with desert, where agents experience gains and losses relative to a reference point determined by what they perceives as their ``deserved'' share of total rewards. \cite{Fallucchi-Trevisan:2024} study a related model for multi-player winner-take-all Tullock contests. \cite{Lim:2010} considers a social reference point based on the modal payoff in the group, akin to inequity aversion \`{a} la \cite{Fehr-Schmidt:1999}. Our setting is closer to \cite{Dato-et-al:2018} and \cite{Fu-et-al:2022}, who consider loss aversion with expectation-based reference points following \cite{kHoszegi-Rabin:2007}. \cite{Dato-et-al:2018} study two-player tournaments with a fixed prize scheme and compare the implications of different personal equilibrium concepts, whereas \cite{Fu-et-al:2022} study winner-take-all Tullock contests with heterogeneous players. Neither paper addresses the optimal allocation of prizes.

The closest studies of prize design with expectation-based loss aversion are \cite{Fu-et-al:2021} and \cite{Ryvkin-Wu:2024}. \cite{Fu-et-al:2021} consider multi-prize nested Tullock contests and show that optimal prizes become more equitable as loss aversion increases. \cite{Ryvkin-Wu:2024} obtain a similar equalization result for rank-order tournaments with general distributions of additive performance shocks. They also show that effort can increase with loss aversion under sufficiently equitable prizes, and that this effect can arise under optimal prizes when performance noise has sufficiently heavy tails. In the present paper, the model is different: there is no noise, rankings are determined directly by effort, and there is ambiguity about convex effort costs. Notably, the previous studies are restricted to relatively weak loss aversion and do not consider the interaction of prize design with endogenous participation.\footnote{\cite{Ryvkin-Wu:2024} show, using an example, that endogenous participation can change their prize equalization result; it is sometimes optimal to sacrifice participation by making prizes more unequal.} 
Equalization under strong loss aversion takes the form of a rising bottom prize, while the number of agents receiving the common top prize remains unchanged. Moreover, maximal robust expected effort decreases with loss aversion, and the equilibrium attaining this value involves declining participation.

Our analysis also requires a modification of the standard \emph{choice-acclimating personal equilibrium} (CPE) concept of \cite{kHoszegi-Rabin:2007} when it comes to the timing of reference-point acclimation to lotteries. As noted by \cite{Dato-et-al:2017}, when expectations acclimate to a lottery over actions, utility is generally nonlinear in the randomization probabilities. We instead assume that reference expectations acclimate to the \emph{realized} effort choice, so utility from a mixed strategy is the expectation of the corresponding choice-acclimated utilities. This approach is related to the one adopted by \cite{Meisner-VonWangenheim:2023} in the context of school choice. It allows us to study symmetric mixed equilibria in all-pay contests while retaining endogenous, expectation-based reference points. Under strong loss aversion, these equilibria can involve nonparticipation, atoms, and gaps in expenditure support. Our robust design argument accommodates these possibilities by establishing an upper bound across all existing symmetric equilibria and constructing an equilibrium that attains it.

Our results are also related to the literature on incentive provision under loss aversion. \cite{Herweg-et-al:2010} study a moral hazard problem and show that a bonus scheme with two payment levels is optimal when the agent is loss averse but risk neutral. Their result and ours both illustrate how coarse reward structures can be optimal under loss aversion, although the contracting environments differ. In our setting, rewards depend on relative performance, and the principal chooses both the prize differential and the bottom prize under a fixed budget. Relatedly, prize allocation has been studied in tournaments with risk-averse but loss-neutral agents \citep{Krishna-Morgan:1998,Kalra-Shi:2001,Akerlof-Holden:2012,Drugov-Ryvkin:2022}. Prize sharing can provide insurance in those settings. Here, agents are risk neutral over monetary payoffs, but prize differences generate psychological losses from unfavorable comparisons to alternative prize states. A positive bottom prize additionally raises the monetary benefit from participation.

Finally, our paper contributes to the literature on robust incentive design \citep[for a review, see][]{Carroll:2019}. \cite{Carroll:2015} shows how a simple linear contract can be optimal when a principal evaluates contracts under the least favorable technology consistent with limited knowledge of the agent's available actions. \cite{Drugov-Ryvkin:2025robust} study robust prize design in rank-order tournaments when the distribution of performance noise is unknown. In the present paper, there is ambiguity about the agents' convex effort costs, subject to a lower bound on productivity under a benchmark expenditure distribution. Transforming effort into cost expenditure separates equilibrium behavior from the unknown cost function and allows us to identify a linear cost function that is least favorable at the optimal schedule. The resulting robust prize schedule, therefore, requires only limited information about effort costs.

\bigskip

The rest of the paper is structured as follows. In Section \ref{sec:model} we set up the model, define our equilibrium concept, and formulate the robust design problem. Our main results are stated in Section \ref{sec:main_result}. The remaining sections present the proof of the main results organized as a sequence of lemmas. We first solve the optimal prize design problem in expenditure space, starting with the moderate loss aversion case in Section \ref{sec:moderate}, and then  considering strong loss aversion with partial and full participation in Section \ref{sec:strong}. We then consider the robust design problem in the effort space in Section \ref{sec:robust}, and Section \ref{sec:conclusions} concludes. All missing proofs are relegated to Appendix \ref{app_proofs}.

\section{The model}
\label{sec:model}

There are $n\ge 2$ identical agents, indexed by $i=1,\ldots,n$. Each agent first decides whether to participate in the contest, and nonparticipants receive zero utility. Conditional on participation, agent $i$ exerts effort $e_i\in\mathds{R}_+$ at cost $c(e_i)$, where $c(\cdot)$ is continuous, strictly increasing, and convex, with $c(0)=0$. Let $x_i=c(e_i)$ denote the agent's cost expenditure. Since $c$ is strictly increasing, ranking agents by effort $e_i$ is equivalent to ranking them by cost expenditure $x_i$. As we observe below, it is convenient to cast the equilibrium analysis in the expenditure space. Effort can then be recovered from cost expenditure as $e_i=q(x_i)$, where $q(\cdot)=c^{-1}(\cdot)$ is the inverse cost function.

The contest pay scheme is a vector of positive, decreasing\footnote{Throughout the paper, the terms ``positive,'' ``negative,'' ``increasing,'' and ``decreasing''  are understood in the weak sense; ``strictly'' is added when necessary.} prizes ${\bf v}=(v_1,\ldots,v_n)$ that satisfy the budget constraint $\sum_{r=1}^nv_r=1$ (where the unit budget is a normalization). We assume that $v_1>v_n$, i.e., the prizes are not all the same. Let $\mathcal{V}$ denote the set of such prize schedules. The agent with the highest effort receives $v_1$, the agent with the second highest effort receives $v_2$, etc. Ties are broken equiprobably.

The agents are loss averse, with endogenous, expectation-based reference points formed following the approach of \cite{kHoszegi-Rabin:2006,kHoszegi-Rabin:2007}. As an equilibrium concept, we adopt a (modified) version of the \emph{choice-acclimating personal equilibrium} (CPE), whereby it is assumed that the agents have correct beliefs about how their efforts affect the probabilities of future outcomes. Following the literature on all-pay contests \citep[e.g.,][]{Barut-Kovenock:1998,Fang-et-al:2020}, we focus on a symmetric equilibrium in mixed strategies.\footnote{Since the equilibrium is in mixed strategies, our version of CPE needs to be carefully defined, to avoid the issues identified by \cite{Dato-et-al:2017}. As explained below, we treat effort realizations, not lotteries over effort, as the choice to which the agents' expectations acclimate. This approach is similar to \cite{Meisner-VonWangenheim:2023} in application to school choice.}

Formally, suppose all agents except one choose cost expenditure according to a mixed strategy with distribution $F(\cdot)$. Consider an agent with cost expenditure $x\ge 0$, and let $p_r(x;F)$ denote the probability for the agent's effort to be ranked $r$. Cost expenditure $x$ thus induces a lottery ${\bf p}(x;F)$ over prizes. Let $U({\bf p}|\tilde{\bf p})$ denote the agent's utility from lottery ${\bf p}$ given a reference lottery $\tilde{\bf p}$. Choice acclimation implies that when the agent chooses $x$, both her consumption lottery and reference lottery are ${\bf p}(x;F)$; therefore, her utility is $u(x;F) = U({\bf p}(x;F)|{\bf p}(x;F))$.

Explicitly, we write the agent's utility from cost expenditure $x$ in the form
\begin{align}
\label{utility}
u(x;F) = \sum_{r=1}^np_r(x;F)v_r - x + \eta\sum_{r=1}^np_r(x;F)\sum_{s\ne r}p_s(x;F)\mu(v_r-v_s).
\end{align}
The first two terms---the expected benefit and cost of effort---represent the monetary utility, and the last term is the psychological utility from gains and losses relative to an expectation-based reference point. Parameter $\eta>0$ is the relative weight of this psychological component, and gain-loss function $\mu(\cdot)$ is defined as
\begin{align}
\label{gain_loss_utility}
\mu(v) = \left\{\begin{array}{ll}v,&v\ge 0\\ \lambda v,& v<0\end{array}\right.
\end{align}
Parameter $\lambda\ge 1$ is the loss aversion coefficient. The gain-loss utility is formed by considering each potential outcome---prize $v_r$ with probability $p_r$---and calculating expected gains and losses relative to all alternative outcomes---prizes $v_s$ with probabilities $p_s$, $s\ne r$---in that state.

Using (\ref{gain_loss_utility}), utility (\ref{utility}) can be simplified as
\begin{align}
\label{utility_loss}
u(x;F) = \sum_{r=1}^np_r(x;F)v_r - x - \theta\sum_{r=1}^np_r(x;F)\sum_{s<r}p_s(x;F)(v_s-v_r),
\end{align}
where we defined parameter $\theta=\eta(\lambda-1)\ge 0$. For $\lambda=1$, gains and losses enter symmetrically into the third term in (\ref{utility}) and cancel each other out. The remaining psychological (dis-)utility is due to losses from higher alternative prizes ($s<r$ in each state with prize $v_r$) in excess of gains when $\lambda>1$.

We look for a symmetric, mixed strategy equilibrium in which all agents choose cost expenditure according to a distribution $F$. We assume that randomization happens before reference point acclimation: An agent draws $x$ from her mixed strategy and implements the corresponding effort $q(x)$; then, conditional on the realized action, reference expectations acclimate to the induced lottery ${\bf p}(x;F)$. Thus, if an agent randomizes cost expenditure with some distribution $G$, we define her utility from such randomization as $u(G;F)=\int u(x;F)\dd G(x)$, with $u$ given by (\ref{utility_loss}).\footnote{For every $x$, the agent evaluates the induced prize lottery relative to the reference lottery generated by $x$. Thus, reference expectations acclimate to the \emph{realized} action, rather than to the \emph{ex ante} lottery over actions. When agents randomize, utility is the expectation of these choice-acclimated utilities across realized actions and therefore has the usual expected-utility form, linear in $G$. Under the standard CPE approach, in contrast, reference expectations would acclimate to the compound lottery generated by $G$, yielding utility $U(\int{\bf p}(x;F)\dd G(x)|\int{\bf p}(x;F)\dd G(x))$, which is nonlinear in $G$.} Our equilibrium concept---\emph{realization-acclimating mixed CPE} (which we hereafter refer to as ``mixed CPE'' for brevity)---is then defined formally as follows.

\begin{definition}
\label{def_equilibrium}
A symmetric \emph{realization-acclimating mixed CPE} is a distribution $F^*$ such that $u(x;F^*)=\bar{u}$ for all $x\in{\rm supp}(F^*)$ and $u(x;F^*)\le\bar{u}$ for all $x\notin{\rm supp}(F^*)$, for some $\bar{u}$.
\end{definition}

It is convenient to proceed by introducing $P_r(x;F)=\sum_{s=1}^rp_s(x;F)$---the probability for the indicative agent to be ranked $r$ \emph{or better}. Additionally, let $d_r=v_r-v_{r+1}$, $r=1,\ldots,n-1$, denote the prize differences, and set $d_n=v_n$. The budget constraint becomes $\sum_{r=1}^nrd_r=1$. Utility (\ref{utility_loss}) can then be transformed as
\begin{align}
\label{utility_P}
u(x;F) = \sum_{r=1}^nP_r(x;F)d_r - x - \theta\sum_{r=1}^nP_r(x;F)[1-P_r(x;F)]d_r,
\end{align}
which we can rewrite as
\[
u(x;F) = B(x;F) - x,
\]
where the benefit function is
\begin{align}
\label{benefit}
B(x;F) = \sum_{r=1}^n[(1-\theta)P_r(x;F)+\theta P_r(x;F)^2]d_r.
\end{align}

Suppose $F(\cdot)$ is an atomless distribution. Then the probability for an agent with cost expenditure $x$ to be ranked $r$, assuming all other agents randomize according to $F$, is $p_r(x;F)=\phi_r(F(x))$, where
\[
\phi_r(q) = \binom{n-1}{r-1}q^{n-r}(1-q)^{r-1}, \quad r=1,\ldots,n, \quad q\in[0,1].
\]
The corresponding cumulative probability of being ranked $r$ or better is $P_r(x;F)=\Phi_r(F(x))$, where $\Phi_r(q)=\sum_{s=1}^r\phi_s(q)$. Note that, for $r=1,\ldots,n-1$, $\Phi_r(q)$ is the cdf of ${\rm Beta}(n-r,r)$,\footnote{Indeed, $\Phi_r(q)=I_q(n-r,r)$, where $I_q(a,b)=\frac{1}{B(n-r,r)}\int_0^qt^{n-r-1}(1-t)^{r-1}\dd t$ is the regularized incomplete beta function.} with pdf
\[
\Phi'_r(q) = \frac{1}{B(n-r,r)}q^{n-r-1}(1-q)^{r-1}.
\]
For $r=n$, $\Phi_r(q)=1$ for all $q\in[0,1]$.

Finally, let
\begin{align}
\label{H_def}
H(q;{\bf v}) = \sum_{r=1}^n\psi(\Phi_r(q))d_r, \quad \text{where }\psi(P) = (1-\theta)P+\theta P^2,
\end{align}
denote the corresponding benefit function, such that $B(x;F)=H(F(x);{\bf v})$. As seen from (\ref{H_def}), $H(q;{\bf v})$ is positive and strictly increasing in $q$ for $\theta\le 1$; however, for $\theta>1$ it can be negative and nonmonotone in $q$. Therefore, in what follows we distinguish two cases: (i) moderate loss aversion, $\theta\le 1$; and (ii) strong loss aversion, $\theta>1$.

The transformation from effort to cost expenditure separates equilibrium from the shape of the cost function. For any strictly increasing $c$, a symmetric mixed CPE in effort maps one-to-one into a symmetric mixed CPE in cost expenditure. Let $\mathcal E({\bf v})$ denote the set of symmetric mixed CPE outcomes in the expenditure space generated by ${\bf v}$. This set depends on ${\bf v}$ and $\theta$, but not on $c$. Let $\mathcal V_E:=\{{\bf v}\in\mathcal V:\mathcal E({\bf v})\ne\varnothing\}$ denote the set of prize schedules that admit at least one symmetric mixed CPE.\footnote{We do not require that $\mathcal V_E=\mathcal V$. Standard equilibrium existence results for discontinuous games and contests with ties \citep[e.g.,][]{Reny:1999,Olszewski-Siegel:2023} are not directly applicable here. Because $\psi(P)$ is decreasing for sufficiently small $P$ under strong loss aversion, a better rank lottery does not necessarily increase utility and the usual tie-breaking argument for establishing better-reply security may fail.}

We assume the principal does not know the agents' cost function, except that it belongs to a certain set. It is convenient to describe the set of admissible cost functions in terms of the inverse cost function $q=c^{-1}$. Since the monetary benefit term in utility (\ref{utility_loss}) does not exceed $v_1\le 1$, we can restrict the domain of $q$ to $[0,1]$. For a given $A>0$, let\footnote{Strict monotonicity ensures that the inverse cost function preserves rankings between effort and cost expenditure. In the minimization below, it will be convenient to use the closure of $\mathcal Q_A$, which also contains \emph{weakly} increasing concave functions. This does not change any infimum because every weakly increasing admissible $q$ can be approximated by strictly increasing admissible functions $q_{\varepsilon}(x)=q(x)+\varepsilon x$.}
\begin{align}
\label{eq:Q_A}
\mathcal Q_A = \left\{q:[0,1]\to\mathds{R}_+: q(0)=0,\ q\text{ strictly increasing and concave},\, \int_0^1q(x)\dd x\ge A
\right\}.
\end{align}
The integral constraint provides a lower bound of overall productivity and rules out arbitrarily high effort costs, while leaving their shape otherwise unrestricted. To interpret this constraint, suppose effort expenditures are distributed as $X_0\sim{\rm Uniform}[0,1]$. Effort associated with expenditure $x$ is $q(x)$, and hence $\mathds{E}[q(X_0)] = \int_0^1 q(x)\dd x$ is the expected effort generated by the benchmark expenditure distribution. Thus, (\ref{eq:Q_A}) requires every admissible technology to generate at least $A$ units of expected effort under this uniform benchmark.\footnote{From an empirical perspective, the restriction can be interpreted as summarizing coarse information about the productivity of effort that a firm may possess from historical variation or internal experiments in incentive schemes and compensation policies. Such data may reveal the average amount of effort generated over a range of incentive intensities, while being insufficient to identify the full shape of the effort cost function. In this sense, constant $A$ captures limited knowledge of the overall productivity of the effort technology.}

For ${\bf v}\in\mathcal V_E$, let $X_\sigma$ denote the equilibrium cost expenditure of a randomly selected potential contestant in equilibrium $\sigma\in\mathcal E({\bf v})$, with $X_\sigma=0$ for a nonparticipant. ``Physical'' effort is $q(X_\sigma)$. For each admissible technology, the principal evaluates a prize schedule by the largest expected effort that can be implemented in a symmetric mixed CPE. Thus, the \emph{robust} value of ${\bf v}$ to the principal is
\begin{align}
\label{eq:robust_value}
W({\bf v}) = \inf_{q\in\mathcal Q_A}
\sup_{\sigma\in\mathcal E({\bf v})}
\mathds{E}\big[q(X_\sigma)\big],
\end{align}
and the principal's problem is
\begin{align}
\label{effort_max}
\max_{{\bf v}\in\mathcal V_E}W({\bf v}).
\end{align}
The inner supremum is taken only over symmetric mixed CPEs that exist. Hence, the robust design problem is well-defined without imposing a general equilibrium existence result for every ${\bf v}\in\mathcal V$.

Problem (\ref{effort_max}) can be intuitively interpreted as a zero-sum game between the principal and Adversary. The principal starts by choosing a prize schedule ${\bf v}\in\mathcal V_E$, Adversary then chooses the worst possible technology $q\in \mathcal Q_A$, and the principal can implement the most productive symmetric mixed CPE under that combination $({\bf v},q(\cdot))$, if one exists.

\section{Optimal prizes: main result}
\label{sec:main_result}

In this section, we state our main result, before formally proving it in the following sections. Under moderate loss aversion, equilibrium is unique and well behaved, similar to the equilibrium of all-pay contests under standard preferences. However, for strong loss aversion, equilibrium may involve nonparticipation, atoms in cost expenditure, gaps in support; and there may also be multiple equilibria. The robust design criterion (\ref{eq:robust_value}) does not require selecting a particular equilibrium across all prize schedules. Instead, for each admissible technology, it evaluates a prize schedule by the most productive symmetric mixed CPE that can be implemented under that technology and prize schedule.

The robustly optimal prize schedule has a very simple form. For any level of loss aversion, it has a ``punish-the-bottom'' structure, with a common prize for ranks $1,\ldots,n-1$ and a lower prize for rank $n$. As loss aversion becomes sufficiently strong, the bottom prize rises and the common top prize falls, making the prize allocation more equal.

Define
\begin{align}
\label{eq:m_star}
m^*(\theta) =
\begin{cases}
0, & \theta\le 1,\\[1mm]
\dfrac{(\theta-1)^2}{4\theta}, & \theta>1.
\end{cases}
\end{align}

\begin{theorem}[Robust optimal prizes under loss aversion]
\label{thm:optimal_prizes_all_theta}
For any $\theta\ge 0$, the prize schedule ${\bf v}^*(\theta)$ given by
\begin{align}
\label{eq:optimal_prizes_main}
v_1^*(\theta)=\cdots=v_{n-1}^*(\theta) =
\frac{1+m^*(\theta)}{n-1+n m^*(\theta)},
\qquad v_n^*(\theta) = \frac{m^*(\theta)}{n-1+n m^*(\theta)}
\end{align}
solves the robust design problem (\ref{effort_max}). Moreover,
\begin{align}
\label{eq:robust_value_star}
W({\bf v}^*(\theta))=2A\,E^*(\theta),
\end{align}
where $E^*(\theta)$ is the maximal expected equilibrium cost expenditure in the normalized linear-cost problem. A least favorable inverse cost function at ${\bf v}^*(\theta)$ is
\begin{align}
\label{eq:q_linear}
q_L(x)=2Ax,
\end{align}
corresponding to the linear cost function $c_L(e)=\frac{e}{2A}$.

For $\theta\in[0,1]$, the optimal schedule supports a unique equilibrium with full participation. For $\theta>1$, the bound is attained by a partial-entry equilibrium, with entry probability
\begin{align}
\label{eq:optimal_entry_main}
\mu^*(\theta) = \left(\frac{\theta+1}{2\theta}\right)^{1/(n-1)}.
\end{align}
\end{theorem}

The \emph{structure} of the robustly optimal prize schedule does not change with loss aversion: it is always a ``punish-the-bottom'' schedule, with a single positive prize differential between ranks $n-1$ and $n$. What changes is the severity of the punishment. For moderate loss aversion, $\theta\le 1$, we have $m^*(\theta)=0$; therefore,
\begin{align}
\label{eq:optimal_moderate_main}
{\bf v}^*(\theta) = \left(
\frac{1}{n-1},\ldots,\frac{1}{n-1},0
\right).
\end{align}
Thus, the standard punish-the-bottom structure with $v_n=0$ remains optimal not only across the range of moderate expectation-based loss aversion, but also under ambiguity about convex effort costs.\footnote{\cite{Fang-et-al:2020} show the optimality of (\ref{eq:optimal_moderate_main}) under standard preferences ($\theta=0$) for any convex cost function. It can be shown that the optimality of (\ref{eq:optimal_moderate_main}) for any convex cost function extends to $\theta\le 1$ as well, see Proposition \ref{prop_moderate_convex} in Appendix \ref{app_additional}.} Once $\theta>1$, $m^*(\theta)>0$, and it becomes optimal to award a strictly positive prize even to the agent ranked last.

The following two corollaries summarize the properties of robustly optimal prizes.

\begin{corollary}[Robustly optimal prizes become more equal with loss aversion]
\label{cor:optimal_equalization}
The robustly optimal prize allocation becomes weakly more equal as $\theta$ increases, and strictly more equal for $\theta>1$. In particular, the optimal prize differential
\begin{align}
\label{eq:optimal_gap}
d_{n-1}^*(\theta)
=
v_{n-1}^*(\theta)-v_n^*(\theta)
=
\frac{1}{n-1+n m^*(\theta)}
\end{align}
is constant for $\theta\le1$ and strictly decreasing for $\theta>1$. Moreover, $\lim_{\theta\to\infty}{\bf v}^*(\theta) = (\frac{1}{n},\ldots,\frac{1}{n})$.
\end{corollary}

\begin{corollary}[Robust value]
\label{cor:maximal_effort}
Under ${\bf v}^*(\theta)$, the worst-case expected effort under the criterion (\ref{eq:robust_value}) is
\begin{align}
\label{eq:maximal_effort}
W({\bf v}^*(\theta))
=
2A\times
\begin{cases}
\dfrac{2n-1-\theta}{n(2n-1)}, & 0\le\theta\le1,\\[3mm]
\dfrac{2\theta(n-1)^2}{n(2n-1)[n-1+nm(\theta)]}
\left(\dfrac{\theta+1}{2\theta}\right)^{\frac{2n-1}{n-1}}, & \theta>1.
\end{cases}
\end{align}
This value is strictly decreasing with $\theta$, with $\lim_{\theta\to\infty}W({\bf v}^*(\theta)) = \frac{4A(n-1)^2}{n^2(2n-1)}2^{-1/(n-1)}>0$.
\end{corollary}

We prove Theorem \ref{thm:optimal_prizes_all_theta} in the next three sections, separating equilibrium analysis from cost ambiguity. Sections \ref{sec:moderate} and \ref{sec:strong} solve an auxiliary problem in the expenditure space. Because $x=c(e)$ is an order-preserving transformation, the set of symmetric mixed CPE distributions of cost expenditure does not depend on the shape of $c$. In this auxiliary problem, the objective is expected cost expenditure $\mathds{E}[X]$, exactly as in a contest with linear effort costs. The key simplification is that the auxiliary problem in the expenditure space is linear in prizes.

For $\theta\le 1$, $\psi(P)=(1-\theta)P+\theta P^2$ is increasing on $[0,1]$. The benefit from expenditure is, therefore, increasing in an agent's equilibrium quantile, and the symmetric equilibrium has the regular properties of an all-pay contest: all agents participate and the expenditure distribution is atomless. The linearity of the problem then implies that one of the \emph{top}-$r$ prize schedules, ${\bf v}^r=(\frac{1}{r},\ldots,\frac{1}{r},0,\ldots,0)$, with one strictly positive prize differential $d_r$, is optimal. Section \ref{sec:moderate} shows that the contribution of a prize differential to expected cost expenditure is increasing in the rank at which it is placed, and hence the punish-the-bottom schedule (\ref{eq:optimal_moderate_main}) maximizes expected cost expenditure.

For $\theta>1$, the situation changes qualitatively. Function $\psi(P)$ can be written as
\begin{align}
\label{eq:psi_main_quadratic}
\psi(P) = \theta[P-P_m(\theta)]^2-m(\theta),
\qquad P_m(\theta)=\frac{\theta-1}{2\theta},
\qquad m(\theta)=\frac{(\theta-1)^2}{4\theta},
\end{align}
and is now initially decreasing. Sufficiently unfavorable rank lotteries then generate negative psychological utility, giving rise to nonparticipation and, for general prize schedules, atoms and gaps in expenditure support.

On a separating part of an equilibrium, cost expenditure equals the benefit net of equilibrium rent. At an atom, the benefit is evaluated using the rank lottery generated by random tie-breaking. The quadratic form of $\psi$ implies that cost expenditure at such an atom is lower than the average raw benefit $H(q;{\bf v})$ over the corresponding quantile interval. Thus, pooling can only \emph{decrease} total expected cost expenditure. The remaining prize allocation problem is linear and can be reduced to extreme points of the budget set. Section \ref{sec:strong} shows that every symmetric mixed CPE outcome under every implementable prize schedule has expected cost expenditure bounded above by the partial-entry equilibrium under ${\bf v}^*(\theta)$.

Section \ref{sec:robust} then shows that this expenditure bound gives the robust result. For any alternative prize schedule, Adversary can choose the linear inverse cost $q_L(x)=2Ax$, so robust expected effort is no larger than $2A$ times maximal expected cost expenditure. Under ${\bf v}^*(\theta)$, one explicitly known equilibrium attains the expenditure bound, and the linear inverse cost is itself least favorable for that equilibrium. The two bounds, therefore, coincide. The proofs of auxiliary results are collected in Appendix \ref{app_proofs}.

\section{Moderate loss aversion ($\theta\le 1$)}
\label{sec:moderate}

In this section, we solve the auxiliary problem in the expenditure space for moderate loss aversion, $\theta\le1$. Throughout Sections \ref{sec:moderate} and \ref{sec:strong}, the variable $x$ and the random variable $X$ refer to cost expenditure. In this space, utility has the form $B-x$, exactly as under linear effort costs. Effort is recovered only in Section \ref{sec:robust} through the inverse cost function $q$. With moderate loss aversion, the symmetric mixed CPE has the same regular properties as the standard symmetric mixed equilibrium in an all-pay contest.

\begin{lemma}
\label{lemma:eq_properties_moderate}
Under moderate loss aversion, the symmetric mixed CPE expenditure distribution $F^*$ is continuous, atomless, and supported on a bounded interval starting from zero.
\end{lemma}

These properties of $F^*$ lead immediately to equilibrium characterization.

\begin{proposition}
\label{prop_eq_moderate}
Under moderate loss aversion, there is a unique symmetric mixed CPE with full participation. The distribution of expenditure is atomless and supported on $[0,v_1-v_n]$, with
\begin{align}
\label{eq:F_moderate}
H(F^*(x);{\bf v})=x+v_n.
\end{align}
The common equilibrium utility is $v_n$.
\end{proposition}

As seen from (\ref{H_def}), $H(q;{\bf v})-v_n$ is decreasing in $\theta$; therefore, the equilibrium cost-expenditure quantile decreases with $\theta$, and we have the following result.

\begin{proposition}
\label{prop:moderate_theta}
Under moderate loss aversion, expenditure decreases with loss aversion in the sense of first-order stochastic dominance.
\end{proposition}

We now turn to the optimal allocation of prizes. For $r=1,\ldots,n-1$, let
\[
J_r(\theta) = \frac1r\int_0^1\psi(\Phi_r(q))\dd q.
\]
Thus, $J_r(\theta)$ is expected cost expenditure under the top-$r$ prize schedule. More generally, equilibrium characterization gives
\begin{align}
\label{eq:effort_moderate_linear}
\mathds{E}[X({\bf v})] = \sum_{r=1}^{n-1}d_r\int_0^1\psi(\Phi_r(q))\dd q
= \sum_{r=1}^{n-1}rd_rJ_r(\theta).
\end{align}
The bottom prize $v_n$ does not enter directly into equilibrium cost expenditure but absorbs $nv_n$ of the prize budget: $\sum_{r=1}^{n-1}rd_r = 1-nv_n$.

\begin{lemma}
\label{lem:moderate_rank_order}
For $\theta\in[0,1]$,
\[
J_{r+1}(\theta)\ge J_r(\theta),
\qquad r=1,\ldots,n-2,
\]
with strict inequality for $\theta>0$.
\end{lemma}

Lemma \ref{lem:moderate_rank_order} and (\ref{eq:effort_moderate_linear}) immediately give the optimal-prize result.

\begin{proposition}
\label{prop_moderate_linear}
Under moderate loss aversion, the prize schedule ${\bf v}^{n-1}=(\frac{1}{n-1},\ldots,\frac{1}{n-1},0)$ maximizes expected cost expenditure for all $\theta\in[0,1]$. For $\theta>0$, it is the unique optimal prize schedule in $\mathcal V$.
\end{proposition}

For $\theta>0$, moderate loss aversion, therefore, makes the punish-the-bottom prize structure strictly optimal in the expenditure space. The proof is based on maximizing (\ref{eq:effort_moderate_linear}) subject to the budget constraint. First, it is immediate that $v_n^*=0$. Second, each $rd_r$ has the same unit ``price'' but its ``benefit'' in (\ref{eq:effort_moderate_linear}) increases with $r$ due to Lemma \ref{lem:moderate_rank_order}; therefore, $(n-1)d_{n-1}=1$ is optimal.

\section{Strong loss aversion ($\theta>1$)}
\label{sec:strong}

We now consider strong loss aversion, $\theta>1$. In this case, the benefit function (\ref{benefit}) can be nonmonotone in the effective quantile and may become negative. The equilibrium characterization is, therefore, qualitatively different from the moderate loss aversion case. To see this, notice that, for $\theta>1$, $\psi(P)<0$ for all sufficiently small $P>0$. Hence, since the prize schedule is nonconstant,
\[
H(q;{\bf v})-v_n
=
\sum_{r=1}^{n-1}d_r\psi(\Phi_r(q))<0
\]
for all sufficiently small $q>0$. Thus, an equilibrium with full entry and atomless $F$ cannot have support starting from zero: if it did, equilibrium utility would equal $H(0;{\bf v})=v_n$, whereas sufficiently small positive expenditures would yield a strictly lower benefit even before effort cost is subtracted. More generally, strong loss aversion can generate nonparticipation, expenditure atoms, gaps in expenditure support, and multiple symmetric mixed CPEs.

The participation constraint, therefore, becomes critical in this setting. We consider a symmetric entry equilibrium with probability of entry $\mu$. Agents who stay out receive an outside option normalized to zero. Conditional on entry, agents choose cost expenditure without observing the number of other entrants. Since full nonparticipation cannot be an equilibrium, $\mu\in(0,1]$.

\subsection{Partial entry}
\label{sec:partial_entry}

Suppose first that $\mu<1$. Entrants must then be indifferent between entering and staying out and hence obtain zero equilibrium payoff. If all entrants except one choose cost expenditure according to an atomless distribution $F$, an entrant choosing cost expenditure $x$ has the \emph{effective quantile}
\[
z=1-\mu+\mu q,
\qquad q=F(x),
\]
which is the probability that she ranks above any particular opponent. Indeed, an opponent stays out with probability $1-\mu$, and enters and chooses cost expenditure below $x$ with probability $\mu F(x)$. Hence, on an atomless part of the support, the equilibrium indifference condition is
\begin{align}
\label{indiff_strong}
H(1-\mu+\mu F(x);{\bf v})=x.
\end{align}

If an equilibrium with partial entry contains an atom at zero expenditure, the zero-expenditure entrants can equivalently be treated as nonparticipants, from other agents' perspective, because other agents' positive-expenditure rank lotteries and expected expenditures are unaffected. We use this representation below and let $z_0\in(0,1)$ denote the resulting mass of nonparticipants. An entrant deviating to zero expenditure then obtains $H(z_0;{\bf v})$, so equilibrium requires
\begin{align}
\label{eq:entry_inequality}
H(z_0;{\bf v})\le 0.
\end{align}
The inequality is strict if the equilibrium positive-expenditure support starts above zero. In the simplest case, expenditure starts at zero, (\ref{eq:entry_inequality}) binds, $z_0$ is the largest root of $H(z;{\bf v})=0$, and $F$ is atomless. We call this type of equilibrium \emph{regular}.

\begin{definition}
A \emph{regular partial-entry equilibrium} is a pair $(\mu,F)$ such that $\mu=1-z_0$, $H(z_0;{\bf v})=0$, function $H(\cdot;{\bf v})$ is strictly increasing on $[z_0,1]$, and $F$ is atomless, supported on an interval starting at zero, and satisfies (\ref{indiff_strong}) on its support.
\end{definition}

As a generalization of top-$r$ schedules introduced in Section \ref{sec:main_result}, consider prize schedules with two prize levels where the bottom prize may be strictly positive. For $r\in\{1,\ldots,n-1\}$ and $t\ge0$, let ${\bf v}^{r,t}$ denote the two-prize schedule
\begin{align}
\label{eq:two_prize_schedule}
v_1^{r,t}=\cdots=v_r^{r,t} = \frac{1+t}{r+nt},
\qquad
v_{r+1}^{r,t}=\cdots=v_n^{r,t}=\frac{t}{r+nt}.
\end{align}
Its benefit function is
\begin{align}
\label{eq:H_rt}
H_{r,t}(z) = H(z;{\bf v}^{r,t}) = \frac{\psi(\Phi_r(z))+t}{r+nt}.
\end{align}
Using (\ref{eq:psi_main_quadratic}), we observe that, assuming $t\le m(\theta)$, the largest root of $H_{r,t}(z)=0$ is $z=\zeta_{r,t}$ that solves 
\begin{align}
\label{P_t}
\Phi_r(\zeta_{r,t}) = P_t(\theta) = P_m(\theta) + \sqrt{\frac{m(\theta)-t}{\theta}}.
\end{align}
Thus, $H_{r,t}(z)$ is strictly increasing on $[\zeta_{r,t},1]$, i.e., prize schedules ${\bf v}^{r,t}$ support regular partial-entry equilibria.

\begin{lemma}
\label{lemma_two-prize}
For every $r\in\{1,\ldots,n-1\}$ and $t\in[0,m(\theta)]$, ${\bf v}^{r,t}$ supports a regular partial-entry equilibrium with cutoff $\zeta_{r,t}$.
\end{lemma}

Let
\begin{align}
\label{eq:E_rt}
E_{r,t}:=\int_{\zeta_{r,t}}^1H_{r,t}(z)\dd z
=\frac{1}{r+nt}\int_{\zeta_{r,t}}^1[\psi(\Phi_r(z))+t]\dd z
\end{align}
denote expected cost expenditure in this regular equilibrium. The following two lemmas show that expected cost expenditure rises when the positive prize differential is moved toward lower ranks and, for rank $n-1$, when the bottom-prize parameter rises toward $m(\theta)$.

\begin{lemma}[$E_{r,t}$ increases in $r$]
\label{lem:r_dominance}
For all $t\in[0,m(\theta)]$ and all $r=1,\ldots,n-2$,
\begin{align}
\label{eq:r_dominance}
E_{r+1,t}\ge E_{r,t}.
\end{align}
\end{lemma}

\begin{lemma}[$E_{n-1,t}$ increases in $t$]
\label{lem:t_dominance}
For all $t\in[0,m(\theta)]$,
\begin{align}
\label{eq:t_dominance}
E_{n-1,t}\le E_{n-1,m(\theta)}.
\end{align}
\end{lemma}

We next turn to arbitrary partial-entry CPEs. Such equilibria may contain expenditure atoms and gaps in expenditure support. The effective quantile interval $[z_0,1]$ can be represented as a union of separating regions and pooling intervals. On a separating region, equilibrium cost expenditure equals $H(z;{\bf v})$. If all effective quantiles $z\in[a,b]$ pool at the same expenditure, let
\[
\bar\Phi_r(a,b)
:=
\frac{1}{b-a}\int_a^b\Phi_r(z)\dd z.
\]
Random equiprobable tie-breaking assigns an agent at the atom an effective quantile uniformly within $[a,b]$, so the probability of being ranked $r$ or higher is $\bar\Phi_r(a,b)$. The corresponding benefit, and hence equilibrium expenditure under partial entry, is
\begin{align}
\label{eq:block_H}
\bar H(a,b;{\bf v})
=
v_n+\sum_{r=1}^{n-1}d_r\psi(\bar\Phi_r(a,b)).
\end{align}
Gaps in expenditure support correspond to upward jumps between the expenditure levels associated with adjacent separating or pooling regions.

Because function $\psi(P)$ is convex, Jensen's inequality implies that pooling always \emph{lowers} total expected cost expenditure relative to the raw benefit function, regardless of how the equilibrium is ironed.

\begin{lemma}[Pooling lowers total expected cost expenditure]
\label{lem:pooling_total}
Consider any partial-entry symmetric mixed CPE generated by ${\bf v}$, and let $z_0$ denote the mass of nonparticipants (including zero-expenditure entrants reclassified as nonparticipants). Then
\begin{align}
\label{eq:pooling_total}
\mathds{E}[X({\bf v})]
\le
\int_{z_0}^1H(z;{\bf v})\dd z.
\end{align}
\end{lemma}
Indeed, $\mathds{E}[X({\bf v})]$ can be written as a sum over separating and pooling intervals. On each separating interval, expenditure equals $H(z;{\bf v})$, and on each pooling interval $[a,b]$ the corresponding mass point contributes $(b-a)\bar{H}(a,b;{\bf v})\le \int_a^b H(z;{\bf v})\dd z$ due to Jensen.  

Importantly, (\ref{eq:pooling_total}) holds with equality for \emph{regular} equilibria. Therefore, if we can show that the right-hand side of (\ref{eq:pooling_total}) is maximized by a prize schedule that generates a regular equilibrium, the optimality of that prize schedule across all partial-entry equilibria will be established. To this end, define the budget shares $\alpha_r=rd_r\ge 0$, $r=1,\ldots,n$, such that $\sum_{r=1}^n\alpha_r=1$. Define also $h_r(z)=\frac{\psi(\Phi_r(z))}{r}$, $r=1,\ldots,n-1$ and $h_n(z)=\frac{1}{n}$. Then $H(z;{\bf v})=\sum_{r=1}^n\alpha_rh_r(z)$; therefore, for a fixed $z_0$, the upper bound in (\ref{eq:pooling_total}) is a linear function of ${\boldsymbol\alpha}$, while entry requires the linear inequality $\sum_{r=1}^n\alpha_rh_r(z_0)\le 0$.

Thus, the set of prize schedules satisfying the entry condition at a given $z_0$ is an intersection of the unit simplex with a halfspace. Its extreme points, therefore, have a simple structure. 

\begin{lemma}[Extreme points at a fixed entry cutoff]
\label{lem:fixed_cutoff_extreme}
Let
\begin{align}
\label{eq:A_polytope}
\mathcal A(z_0) = \left\{
{\boldsymbol\alpha}\in\mathds{R}_+^n:
\sum\nolimits_{r=1}^n\alpha_r=1, \quad
\sum\nolimits_{r=1}^n\alpha_rh_r(z_0)\le 0
\right\}.
\end{align}
Every extreme point of $\mathcal A(z_0)$ has at most two positive components. If two components are positive, the entry constraint binds.
\end{lemma}

The extreme points of $\mathcal A(z_0)$ can be of three types. An extreme point with one positive component $\alpha_r>0$, where $r\le n-1$, is a top-$r$ schedule with zero bottom prize. An extreme point with two positive components $\alpha_r,\alpha_s>0$, with $r,s\le n-1$, is a convex combination of two such prize schedules. Finally, an extreme point with a positive bottom prize and one other positive prize differential is a two-prize schedule ${\bf v}^{r,t}$ with $t\in[0,m(\theta)]$. In all three cases, Lemmas \ref{lem:r_dominance} and \ref{lem:t_dominance} bound the right-hand side of (\ref{eq:pooling_total}) by $E_{n-1,m(\theta)}$.

\begin{lemma}[Every fixed-cutoff extreme point is dominated by ${\bf v}^{n-1,m(\theta)}$]
\label{lem:extreme_point_dom}
For any extreme point ${\boldsymbol\alpha}$ of $\mathcal A(z_0)$,
\begin{align}
\label{eq:extreme_dom}
\int_{z_0}^1H(z;{\bf v})\dd z
\le
E_{n-1,m(\theta)}.
\end{align}
\end{lemma}

Combining Lemmas \ref{lem:pooling_total}--\ref{lem:extreme_point_dom} gives the optimality result under strong loss aversion for partial entry.

\begin{proposition}
\label{prop:universal_two_prize_linear}
For any ${\bf v}\in\mathcal V$ and any partial-entry symmetric mixed CPE generated by ${\bf v}$,
\begin{align}
\label{eq:partial_universal_bound}
\mathds{E}[X({\bf v})]
\le
E_{n-1,m(\theta)}.
\end{align}
Equality is attained by the regular partial-entry equilibrium under ${\bf v}^{n-1,m(\theta)}$.
\end{proposition}

\paragraph{Entry under optimal prizes}
Consider the partial-entry equilibrium under ${\bf v}^{n-1,m(\theta)}$. We have
\begin{align}
\label{eq:H_optimal_main}
H(z;{\bf v}^{n-1,m(\theta)}) = \frac{
m(\theta)+\psi(\Phi_{n-1}(z))}{n-1+nm(\theta)}
= \frac{\theta[\Phi_{n-1}(z)-P_m(\theta)]^2}{n-1+nm(\theta)}.
\end{align}
Thus, the partial-entry cutoff $z_0$ satisfies $\Phi_{n-1}(z_0)=P_m(\theta)$. Since $\Phi_{n-1}(z)=1-(1-z)^{n-1}$, we obtain entry probability
\begin{align}
\label{eq:mu_star_strong}
\mu = 1-z_0 = \left(\frac{\theta+1}{2\theta}\right)^{1/(n-1)}.
\end{align}
We observe that the optimal benefit function $H(z;{\bf v}^{n-1,m(\theta)})$ is tangent to zero at $z_0$, and is strictly increasing for $z>z_0$, producing a regular equilibrium. The probability of entry declines with loss aversion and increases in the number of agents. 

\subsection{Full entry}
\label{subsec:full_entry}

We now show that allowing for full entry does not change the inequality in Proposition \ref{prop:universal_two_prize_linear}. This result is again independent of equilibrium selection, even though a full-entry CPE may include expenditure atoms, a strictly positive minimum expenditure, or gaps in expenditure support.

The first observation is that strictly positive equilibrium rent is never optimal for expenditure maximization.

\begin{lemma}[Positive rent is dominated]
\label{lem:positive_rent_dominated}
Consider a nondegenerate full-entry symmetric mixed CPE under ${\bf v}$, in which agents get common utility $u>0$. Then there exists a feasible prize schedule $\tilde{\bf v}$ supporting a full-entry symmetric mixed CPE with zero rent in which cost expenditure is weakly higher state by state.
\end{lemma}
As we show in the proof, positive equilibrium rent can be removed via an affine transformation of the prize schedule. Subtracting the same amount from every prize reduces agents' monetary payoff by that amount but leaves all prize \emph{differences}, and hence the gain-loss utility, unchanged. The remaining prizes can then be rescaled to fit the budget, which strengthens incentives proportionally. The corresponding increasing affine transformation of equilibrium expenditures preserves all rankings and ties, reduces rent to zero, and weakly raises expenditures in all states. 

We can, therefore, restrict attention to full-entry equilibria with zero rent. If such an equilibrium has an atom at zero expenditure, this atom can be reinterpreted as nonparticipation. The distribution of positive expenditures and expected expenditure are unchanged, and the resulting partial-entry CPE is covered by Proposition \ref{prop:universal_two_prize_linear}. If there is no atom at zero, zero rent implies $v_n=0$. The same pooling argument as in Lemma \ref{lem:pooling_total} then gives
\[
\mathds{E}[X({\bf v})] \le \int_0^1H(z;{\bf v})\dd z.
\]
Budget shares $\alpha_r=rd_r$ now satisfy $\sum_{r=1}^{n-1}\alpha_r=1$ and
\[
\int_0^1H(z;{\bf v})\dd z = \sum_{r=1}^{n-1}\alpha_r\int_0^1h_r(z)\dd z.
\]
For each $r$, $h_r(z)$ is negative before its positive root and positive thereafter; therefore, the integral in the right-hand side is bounded above by $E_{r,0}$, cf. (\ref{eq:E_rt}). Lemmas \ref{lem:r_dominance} and \ref{lem:t_dominance} then imply that the right-hand side is bounded by $E_{n-1,m(\theta)}$.

Combining these observations with Proposition \ref{prop:universal_two_prize_linear}, we obtain the general optimality result in the strong loss aversion case.

\begin{corollary}[Expenditure optimum under strong loss aversion]
\label{cor:unrestricted_optimum_strong}
Among all symmetric mixed CPE outcomes, expected cost expenditure is maximized by the regular partial-entry equilibrium with ${\bf v}={\bf v}^{\,n-1,m(\theta)}$.
\end{corollary}

Let $E^*(\theta)$ denote the resulting maximal expected cost expenditure. From Proposition \ref{prop_moderate_linear} and Corollary \ref{cor:unrestricted_optimum_strong},
\begin{align}
\label{eq:E_star}
E^*(\theta) = \begin{cases}
\dfrac{2n-1-\theta}{n(2n-1)}, & 0\le\theta\le1,\\
\dfrac{2\theta(n-1)^2}{n(2n-1)[n-1+nm(\theta)]}
\left(\dfrac{\theta+1}{2\theta}\right)^{\frac{2n-1}{n-1}}, & \theta>1.
\end{cases}
\end{align}
This expression gives the following result.

\begin{lemma}[Maximal expected expenditure and loss aversion]
\label{lem:optimal_effort_theta}
Maximal expected expenditure $E^*(\theta)$ is continuous and strictly decreasing in $\theta$, with the limit
\begin{equation}
\label{eq:optimal_effort_limit}
\lim_{\theta\to\infty} E^*(\theta) = \frac{2^{\frac{n-2}{n-1}}(n-1)^2}{n^2(2n-1)} > 0.
\end{equation}
\end{lemma}

The positive limit of $E^*(\theta)$ in Lemma \ref{lem:optimal_effort_theta} may appear surprising because the optimal prize allocation converges to equal prize sharing as $\theta\to\infty$. The reason is that the vanishing prize differential is offset by increasing loss aversion. Indeed,
\[
d_{n-1}^*(\theta) = \frac{1}{n-1+nm(\theta)} \sim \frac{4}{n\theta},
\]
which gives
\[
\theta d_{n-1}^*(\theta)\to\frac{4}{n}.
\]
Thus, although the monetary incentives associated with punishment at the bottom vanish, the loss-aversion component of incentives remains positive, converging to
\[
H(z;{\bf v}^*(\theta)) = \frac{\theta[\Phi_{n-1}(z)-P_m(\theta)]^2}
{n-1+nm(\theta)} \to \frac{4}{n}\left[\Phi_{n-1}(z)-\frac{1}{2}\right]^2.
\]

\section{Robust contest design under cost uncertainty}
\label{sec:robust}

We now return to effort and solve the robust contest design problem (\ref{effort_max}). First, we characterize the least favorable inverse cost for an arbitrary distribution of cost expenditure.

\begin{lemma}[Worst-case inverse cost]
\label{lem:worst_inverse_cost}
Let $X$ be any random variable supported on $[0,1]$. Then
\begin{align}
\label{eq:worst_inverse_cost}
\inf_{q\in\mathcal Q_A}\mathds{E}[q(X)] = A\inf_{t\in(0,1]}
\frac{\mathds{E}[\min\{X,t\}]}{t-\frac{t^2}{2}}.
\end{align}
In particular, for every $q\in\mathcal Q_A$ and every $x\in[0,\frac{1}{2}]$,
\begin{align}
\label{eq:pointwise_linear_bound}
q(x)\ge 2Ax.
\end{align}
\end{lemma}

To understand this result, recall that $\mathcal Q_A$ is the set of increasing, concave inverse cost functions satisfying the productivity constraint
\begin{align}
\label{productivity_constraint}
\int_0^1 q(x)\dd x\ge A.
\end{align}
Any weakly increasing concave function in the closure of $\mathcal Q_A$ can be represented as a linear combination with nonnegative weights, over $t$, of truncated linear functions of the form $\min\{x,t\}$. For such a function, expected effort generated by expenditure $X$ is $\mathds{E}[\min\{X,t\}]$, while its productivity is
\[
\int_0^1\min\{x,t\}\dd x = t-\frac{t^2}{2}.
\]
Thus, the normalized truncated linear technology in the closure of $\mathcal Q_A$ that satisfies (\ref{productivity_constraint}) with equality is
\[
q_t(x) = \frac{A\min\{x,t\}}{t-\frac{t^2}{2}}.
\]
Since both the objective, $\mathds{E}[q(X)]$, and constraint (\ref{productivity_constraint}) are linear in the mixture weights, the least favorable technology in the closure concentrates all weight on the truncation $t$ that generates the smallest expected effort relative to its benchmark productivity, or approaches the infimum along a minimizing sequence of truncations, giving (\ref{eq:worst_inverse_cost}). Since $\mathcal Q_A$ is dense in its closure, the same infimum obtains over $\mathcal Q_A$. Thus, when the infimum is attained, a least favorable technology in the closure is a \emph{truncated linear technology}; otherwise, the infimum can be approached by such technologies. A truncated linear technology is linear in cost expenditure up to some threshold and becomes infinitely unproductive beyond it. The pointwise bound (\ref{eq:pointwise_linear_bound}) follows because $q_t(x)\ge 2Ax$ for all $x\le\frac{1}{2}$.

Second, we show that the linear inverse cost is least favorable for the distribution of expenditure in an equilibrium that attains the expenditure bound (\ref{eq:E_star}).

\begin{lemma}[Linear costs are least favorable under optimal prizes]
\label{lem:linear_least_favorable}
Let $X^*(\theta)$ denote cost expenditure in the equilibrium attaining $E^*(\theta)$ under ${\bf v}^*(\theta)$. Then
\begin{align}
\label{eq:least_favorable_candidate}
\inf_{q\in\mathcal Q_A}\mathds{E}[q(X^*(\theta))] = 2A\mathds{E} [X^*(\theta)] = 2AE^*(\theta).
\end{align}
The infimum is attained by $q_L(x)=2Ax$.
\end{lemma}

For $n\ge 3$, Lemma \ref{lem:linear_least_favorable} follows immediately from (\ref{eq:pointwise_linear_bound}). Indeed, under ${\bf v}^*(\theta)$,
\[
0\le X^*(\theta)\le v_1^*(\theta) \le\frac{1}{n-1}\le\frac{1}{2}.
\]
For $n=2$, equilibrium cost expenditure can exceed $\frac{1}{2}$, and the proof verifies this case separately.

We can now prove Theorem \ref{thm:optimal_prizes_all_theta}. Fix any ${\bf v}\in\mathcal V_E$. Since $q_L(x)=2Ax\in\mathcal Q_A$, we have
\begin{align}
W({\bf v}) = \inf_{q\in\mathcal Q_A}\sup_{\sigma\in\mathcal E({\bf v})}
\mathds{E}[q(X_\sigma)] \le \sup_{\sigma\in\mathcal E({\bf v})}
\mathds{E}[q_L(X_\sigma)] = 2A\sup_{\sigma\in\mathcal E({\bf v})}
\mathds{E}[X_\sigma] \le 2AE^*(\theta),
\label{eq:robust_upper}
\end{align}
where the last inequality follows from Proposition \ref{prop_moderate_linear} for $\theta\le1$ and Corollary \ref{cor:unrestricted_optimum_strong} for $\theta>1$. Notice that this step does not require us to characterize the equilibrium attaining the inner supremum because the expenditure bound applies to every existing symmetric mixed CPE.

Now recall that ${\bf v}^*(\theta)$ supports an explicitly known equilibrium $\sigma^*$ attaining $E^*(\theta)$---the full-entry regular equilibrium described in Section \ref{sec:moderate} for $\theta\le 1$, and the regular partial-entry equilibrium described in Section \ref{sec:partial_entry} for $\theta>1$. Hence, for every $q\in\mathcal Q_A$,
\[
\sup_{\sigma\in\mathcal E({\bf v}^*(\theta))}\mathds{E}[q(X_\sigma)]\ge\mathds{E}[q(X^*)].
\]
Taking the infimum over $q$ and applying Lemma \ref{lem:linear_least_favorable} gives
\begin{align}
W({\bf v}^*(\theta)) \ge \inf_{q\in\mathcal Q_A}\mathds{E}[q(X^*)]
= 2AE^*(\theta).
\label{eq:robust_lower}
\end{align}
Combining (\ref{eq:robust_upper}) and (\ref{eq:robust_lower}) proves Theorem \ref{thm:optimal_prizes_all_theta}, where (\ref{eq:maximal_effort}) follows from (\ref{eq:E_star}).

Thus, our robustness approach largely circumvents the equilibrium characterization and selection problems that arise under strong loss aversion. For an arbitrary prize schedule, we do not need to characterize all symmetric mixed CPEs, determine which equilibrium is selected, or even identify the equilibrium that maximizes effort. The linear cost analysis provides a uniform upper bound, $E^*(\theta)$, on cost expenditure in every symmetric mixed CPE under every feasible prize schedule. Since the linear inverse cost $q_L(x)=2Ax$ belongs to $\mathcal Q_A$, this same bound delivers the robust upper bound for all prize schedules. To complete the argument, it is then sufficient to specify an equilibrium under ${\bf v}^*(\theta)$ that attains the expenditure bound $E^*(\theta)$ and for which the linear technology is least favorable.

\section{Conclusions}
\label{sec:conclusions}

We study robust prize design in all-pay contests with expectation-based loss aversion when the principal does not know the agents' convex cost of effort. For each cost technology, the principal evaluates a prize schedule by the largest expected effort attainable in a symmetric mixed CPE.

In the cost expenditure space, the equilibrium problem is independent of the shape of the cost function. A punish-the-bottom prize schedule maximizes expected expenditure across all symmetric mixed CPE outcomes. The corresponding expenditure bound can then be used for robust design because the least favorable technology at the optimum is linear. 

As loss aversion increases, the robustly optimal allocation of prizes becomes more equal: under moderate loss aversion the agent ranked last receives zero, while under strong loss aversion the bottom prize is positive and increasing in the degree of loss aversion.

Importantly, partial participation is optimal under strong loss aversion. Our results, therefore, suggest that encouraging more agents to enter can be costly even when entry itself is costless: Under a fixed prize budget, making participation more attractive leaves less scope for rewarding relative performance. Our analysis shows that a simple punish-the-bottom prize schedule with a positive bottom prize can achieve the optimal balance between participation and incentives in the presence of uncertainty about effort costs.

\newpage
\bibliographystyle{aea}
\bibliography{Contests_loss_aversion}

\newpage
\appendix

\section{Proofs}
\label{app_proofs}

\begin{proof}{\bf of Lemma \ref{lemma:eq_properties_moderate}}
We show that $F^*$ has no atoms. By contradiction, suppose there is an atom of size $\alpha>0$ at some $x$. Consider increasing cost expenditure from $x$ to $x+\varepsilon$, for $\varepsilon>0$. This strictly raises each $P_r(x;F^*)$ for $r<n$. Because the prize vector is nonconstant, there exists at least one $r<n$ such that $d_r>0$; therefore, $B(x+\varepsilon;F^*)>B(x;F^*)$, and the gain remains strictly positive for $\varepsilon\to0$, whereas the difference in expenditures is $\varepsilon\to0$. Thus, $u(x+\varepsilon;F^*)>u(x;F^*)$ for $\varepsilon>0$ sufficiently small, which contradicts the assumption that $x\in{\rm supp}(F^*)$. The rest of the properties of $F^*$---that the support starts from zero, contains no gaps and is bounded---follow from standard arguments for all-pay contests.
\end{proof}

\bigskip

\begin{proof}{\bf of Proposition \ref{prop_eq_moderate}}
By Lemma \ref{lemma:eq_properties_moderate}, the expenditure distribution is atomless and has connected support $[0,\bar x]$. Since the support starts at zero, $F^*(0)=0$. An agent choosing zero expenditure, therefore, obtains $H(0;{\bf v})=v_n$. Equilibrium indifference implies that the common utility is $v_n$, hence $H(F^*(x);{\bf v})-x=v_n$ throughout the support, which gives (\ref{eq:F_moderate}).

Since $H(\cdot;{\bf v})$ is continuous and strictly increasing, this equation uniquely determines $F^*$. At the upper end of the support, $F^*(\bar x)=1$, so $\bar x=H(1;{\bf v})-H(0;{\bf v})=v_1-v_n$.

For $x>\bar x$, an agent wins the top prize with certainty, which gives $u(x;F^*)=v_1-x<v_n$, so there is no profitable deviation.

The equilibrium utility is $v_n\ge0$. If $v_n>0$, participation is strictly optimal. If $v_n=0$, partial participation cannot occur: any partial-entry cutoff $z_0>0$ would require a nonpositive payoff from entering at zero effort, whereas $H(z_0;{\bf v})>H(0;{\bf v})=0$. Thus, participation is full.
\end{proof}

\bigskip

\begin{proof}{\bf of Proposition \ref{prop:moderate_theta}}
From Proposition \ref{prop_eq_moderate}, the equilibrium expenditure quantile is
\[
x(q)=H(q;{\bf v})-v_n
=\sum_{r=1}^{n-1}d_r\psi(\Phi_r(q)).
\]
For every $P\in[0,1]$,
\[
\frac{\partial\psi(P)}{\partial\theta}
=-P(1-P)\le0.
\]
Thus, $x(q)$ decreases pointwise in $\theta$, proving first-order stochastic dominance.
\end{proof}

\bigskip

\begin{proof}{\bf of Lemma \ref{lem:moderate_rank_order}}
Recall that
\[
J_r(\theta) = \frac{1}{r}\int_0^1\left[(1-\theta)\Phi_r(q)+\theta\Phi_r(q)^2\right]\dd q,
\]
which gives
\begin{align}
\label{J_diff}
& J_{r+1}(\theta) - J_r(\theta) = (1-\theta)\left[\frac{1}{r+1}\int_0^1\Phi_{r+1}(q)\dd q - \frac{1}{r}\int_0^1\Phi_{r}(q)\dd q\right]  \\ 
& + \theta\left[\frac{1}{r+1}\int_0^1\Phi_{r+1}(q)^2\dd q - \frac{1}{r}\int_0^1\Phi_{r}(q)^2\dd q\right].\nonumber 
\end{align}
Note that
\[
\int_0^1\phi_s(q)\dd q = \binom{n-1}{s-1}B(n-s+1,s) = \frac{1}{n},
\]
giving
\[
\frac{1}{r}\int_0^1\Phi_r(q)\dd q = \frac{1}{r}\sum_{s=1}^r\int_0^1\phi_s(q)\dd q = \frac{1}{n},
\]
and hence the difference between the linear terms in (\ref{J_diff}) vanishes. 

For the quadratic terms, writing $\Phi_{r+1}=\Phi_r+\phi_{r+1}$ gives
\begin{align}
\label{J_diff_qudratic}
&\frac{1}{r+1}\int_0^1\Phi_{r+1}(q)^2\dd q -
\frac{1}{r}\int_0^1\Phi_r(q)^2\dd q\nonumber\\
& = -\frac{1}{r(r+1)}\int_0^1\Phi_r(q)^2\dd q + \frac{2}{r+1}
\int_0^1\Phi_r(q)\phi_{r+1}(q)\dd q + \frac{1}{r+1}\int_0^1\phi_{r+1}(q)^2\dd q.
\end{align}
It can be directly verified that
\begin{align}
\label{phi_Phi_identity}
\phi_{r+1}(q) = \frac{1-q}{r}\Phi_r'(q).
\end{align}
Using this identity and integrating by parts, we have
\begin{align*}
& 2\int_0^1\Phi_r(q)\phi_{r+1}(q)\dd q = \frac{2}{r}\int_0^1
(1-q)\Phi_r(q)\Phi_r'(q)\dd q \\
& = \frac{1}{r}\int_0^1(1-q)\frac{\dd}{\dd q}\left[\Phi_r(q)^2\right]\dd q
= \frac{1}{r}\left[(1-q)\Phi_r(q)^2\big|_0^1 + \int_0^1\Phi_r(q)^2\dd q
\right] = \frac{1}{r}\int_0^1\Phi_r(q)^2\dd q.
\end{align*}
Therefore, the first two terms in (\ref{J_diff_qudratic}) cancel out and we obtain
\[
\frac{1}{r+1}\int_0^1\Phi_{r+1}(q)^2\dd q -
\frac{1}{r}\int_0^1\Phi_r(q)^2\dd q = \frac{1}{r+1}\int_0^1\phi_{r+1}(q)^2\dd q,
\]
and hence
\[
J_{r+1}(\theta) - J_r(\theta) = \frac{\theta}{r+1}\int_0^1\phi_{r+1}(q)^2\dd q \ge 0.
\]
The inequality is strict for $\theta>0$.
\end{proof}

\bigskip

\begin{proof}{\bf of Proposition \ref{prop_moderate_linear}}
Equation (\ref{eq:effort_moderate_linear}) gives
\[
\mathds{E}[X({\bf v})] = \sum_{r=1}^{n-1}(rd_r)J_r(\theta).
\]
By Lemma \ref{lem:moderate_rank_order}, $J_r(\theta)\le J_{n-1}(\theta)$ for all $r<n$. From the budget constraint we have
\[
\sum_{r=1}^{n-1}rd_r=1-nv_n\le1,
\]
and $J_{n-1}(\theta)>0$; therefore,
\begin{align}
\label{ineq_moderate}
\mathds{E}[X({\bf v})] \le (1-nv_n)J_{n-1}(\theta) \le J_{n-1}(\theta).
\end{align}
Note that $J_{n-1}(\theta)$ is expected expenditure under ${\bf v}^{n-1}$. For $\theta>0$, the inequality in Lemma \ref{lem:moderate_rank_order} is strict for every $r<n-1$, so (\ref{ineq_moderate}) holds with equality only if $v_n=0$ and $d_r=0$ for $r<n-1$. This implies the uniqueness of the optimum. For $\theta=0$, $J_r(0)=\frac{1}{n}$ for each $r$, so any schedule with $v_n=0$ is optimal.
\end{proof}

\bigskip

\begin{proof}{\bf of Lemma \ref{lemma_two-prize}}
For prize schedule ${\bf v}^{r,t}$, we have
\[
H_{r,t}(z) = H(z;{\bf v}^{r,t}) = \frac{\psi(\Phi_r(z))+t}{r+nt}.
\]
The roots of $H_{r,t}(z)$ are, therefore, determined by $\psi(P)+t=0$. Using (\ref{eq:psi_main_quadratic}), a root exists if and only if $t\le m(\theta)$, and the larger root is $P_t(\theta)$ in (\ref{P_t}). Since $\Phi_r$ is continuous and strictly increasing, the corresponding largest root of $H_{r,t}$ is $\zeta_{r,t}=\Phi_r^{-1}(P_t)$.

Differentiating gives
\[
H'_{r,t}(z)
=
\frac{1}{r+nt}\psi'(\Phi_r(z))\Phi_r'(z).
\]
Function $\psi$ reaches its minimum at $P_m(\theta)=\frac{\theta-1}{2\theta}$, and $P_t(\theta)\ge P_m(\theta)$; therefore, $\psi'(\Phi_r(z))>0$ for $z>\zeta_{r,t}$. Since $\Phi'_r(z)>0$, $H_{r,t}(z)$ is strictly increasing for $z>\zeta_{r,t}$. Hence, ${\bf v}^{r,t}$ supports the regular partial-entry equilibrium.
\end{proof}

\bigskip

\begin{proof}{\bf of Lemma \ref{lem:r_dominance}}
Define $Q_t(P) = \psi(P)+t$. From (\ref{phi_Phi_identity}), we have
\begin{align}
\label{eq:Phi_recursion_linear}
\Phi_{r+1}(z)=\Phi_r(z)+\frac{1-z}{r}\Phi_r'(z).
\end{align}
Using the expression $\psi(P)=(1-\theta)P + \theta P^2$, we then write
\begin{align}
\label{eq:Q_recursion_linear}
& Q_t(\Phi_{r+1}(z)) = (1-\theta)\left[\Phi_r(z)+\frac{1-z}{r}\Phi_r'(z)\right] + \theta\left[\Phi_r(z)+\frac{1-z}{r}\Phi_r'(z)\right]^2 + t \nonumber\\
& = Q_t(\Phi_r(z)) + \frac{1-z}{r}\left[(1-\theta)\Phi_r'(z) + 2\theta \Phi_r(z)\Phi_r'(z)\right] + \frac{\theta(1-z)^2}{r^2}\Phi_r'(z)^2 \nonumber\\
& = Q_t(\Phi_r(z)) + \frac{1-z}{r}\frac{\dd}{\dd z}Q_t(\Phi_r(z)) + \frac{\theta(1-z)^2}{r^2}\Phi_r'(z)^2.
\end{align}
Integrating from $\zeta_{r,t}$ to $1$, with $Q_t(\Phi_r(\zeta_{r,t}))=0$, gives
\begin{align}
\label{eq:Q_integrated_linear}
\int_{\zeta_{r,t}}^1Q_t(\Phi_{r+1}(z))\dd z = \frac{r+1}{r}\int_{\zeta_{r,t}}^1Q_t(\Phi_r(z))\dd z + \frac{\theta}{r^2}\int_{\zeta_{r,t}}^1(1-z)^2\Phi_r'(z)^2\dd z.
\end{align}
Moreover, $\Phi_{r+1}(z)>\Phi_r(z)$ for $z\in(0,1)$, which implies $\zeta_{r+1,t}<\zeta_{r,t}$. Because $Q_t(\Phi_{r+1}(z))\ge 0$ for $z\ge\zeta_{r+1,t}$, we have from (\ref{eq:E_rt})
\begin{align*}
E_{r+1,t} = \frac{1}{r+1+nt}\int_{\zeta_{r+1,t}}^1Q_t(\Phi_{r+1}(z))\dd z \ge \frac{1}{r+1+nt}\int_{\zeta_{r,t}}^1Q_t(\Phi_{r+1}(z))\dd z.
\end{align*}
Combining this inequality with (\ref{eq:Q_integrated_linear}) yields
\begin{align*}
& E_{r+1,t}-E_{r,t} \ge \frac{1}{r+1+nt}\int_{\zeta_{r,t}}^1Q_t(\Phi_{r+1}(z))\dd z - \frac{1}{r+nt}\int_{\zeta_{r,t}}^1Q_t(\Phi_r(z))\dd z \\
& \ge \frac{1}{r+1+nt}\frac{r+1}{r}\int_{\zeta_{r,t}}^1Q_t(\Phi_r(z))\dd z - \frac{1}{r+nt}\int_{\zeta_{r,t}}^1Q_t(\Phi_r(z))\dd z \\
& = \frac{nt}{r(r+nt)(r+1+nt)}\int_{\zeta_{r,t}}^1Q_t(\Phi_r(z))\dd z \ge 0.
\end{align*}
\end{proof}

\bigskip

\begin{proof}{\bf of Lemma \ref{lem:t_dominance}}
For $t\in[0,m(\theta)]$,
\[
E_{n-1,t}
=
\frac{1}{n-1+nt}\int_{\zeta_{n-1,t}}^1[\psi(\Phi_{n-1}(z))+t]\dd z.
\]
Differentiating with respect to $t$ and noting that the integrand is zero at $z=\zeta_{n-1,t}$ gives
\begin{align}
\label{eq:dEdt_linear}
\frac{\partial E_{n-1,t}}{\partial t} = \frac{1}{(n-1+nt)^2}\int_{\zeta_{n-1,t}}^1
[n-1-n\psi(\Phi_{n-1}(z))]\dd z.
\end{align}
Let $u=1-z$, then $\Phi_{n-1}(z)=1-(1-z)^{n-1} = 1-u^{n-1}$, and the integrand in (\ref{eq:dEdt_linear}) is
\[
n-1-n(1-\theta)(1-u^{n-1}) - n\theta(1-u^{n-1})^2 = -1 + n(\theta+1)u^{n-1} - n\theta u^{2n-2}.
\]
Its antiderivative with respect to $u$, therefore, is
\begin{align}
\label{eq:F_u_linear}
F(u)=-u+(1+\theta)u^n-\frac{n\theta}{2n-1}u^{2n-1}.
\end{align}
Thus, the integral in (\ref{eq:dEdt_linear}) equals $F(1-\zeta_{n-1,t})-F(0)$.

Let $y=u^{n-1}$ and write
\[
F(u)=uG(y),
\qquad
G(y)=-1+(1+\theta)y-\frac{n\theta}{2n-1}y^2.
\]
From (\ref{P_t}),
\[
(1-\zeta_{n-1,t})^{n-1} = 1-P_t(\theta) \in \left[\frac1\theta,\frac{\theta+1}{2\theta}\right].
\]
Function $G(y)$ is concave, and both
\[
G\left(\frac1\theta\right) = \frac{n-1}{(2n-1)\theta}>0,
\]
and
\[
G\left(\frac{\theta+1}{2\theta}\right) = -1+\frac{(\theta+1)^2}{4\theta}\frac{3n-2}{2n-1}>0.
\]
Therefore, $F(1-\zeta_{n-1,t})>0=F(0)$, so (\ref{eq:dEdt_linear}) is positive, which proves (\ref{eq:t_dominance}).
\end{proof}

\bigskip

\begin{proof}{\bf of Lemma \ref{lem:pooling_total}}
After zero-expenditure entrants are reclassified as nonparticipants, the effective quantile interval of entrants is $[z_0,1]$. On a separating region, zero equilibrium rent implies that cost expenditure equals $H(z;{\bf v})$. On a pooling interval $[a,b]$, expenditure is $\bar H(a,b;{\bf v})$.

For any random variable $Y$ with support contained in $[0,1]$, the quadratic form of $\psi(P)$ implies
\[
\mathds{E}[\psi(Y)] = \psi(\mathds{E}[Y])+\theta{\rm Var}[Y].
\]
Applying this identity to $Y=\Phi_r(Z)$, where $Z\sim {\rm Uniform}[a,b]$, and summing with weights $d_r$ gives
\begin{align}
\label{eq:variance_block_identity}
\frac{1}{b-a}\int_a^b H(z;{\bf v})\dd z = \bar H(a,b;{\bf v}) +
\theta\sum_{r=1}^{n-1}d_r{\rm Var}[\Phi_r(Z)].
\end{align}
Therefore,
\[
(b-a)\bar H(a,b;{\bf v}) \le \int_a^bH(z;{\bf v})\dd z.
\]
Summing this inequality over all pooling intervals and using equality on separating regions gives
\[
\mathds{E}[X({\bf v})] \le \int_{z_0}^1H(z;{\bf v})\dd z.
\]
\end{proof}

\bigskip

\begin{proof}{\bf of Lemma \ref{lem:fixed_cutoff_extreme}}
Set $\mathcal A(z_0)$ is an intersection of the unit simplex and a halfspace. An extreme point that does not lie on the boundary of the halfspace is, therefore, an extreme point of the simplex and has one positive component. On the boundary $\sum_{r=1}^n\alpha_rh_r(z_0)=0$, there are two independent equality constraints, so an extreme point has at most two positive components.
\end{proof}

\bigskip

\begin{proof}{\bf of Lemma \ref{lem:extreme_point_dom}}
For $r=1,\ldots,n-1$, let $A_r(z_0):=\int_{z_0}^1h_r(z)\dd z$. Let $\zeta_{r,0}$ denote the positive root of $h_r(z)$. Since $h_r(z)$ is negative on $(0,\zeta_{r,0})$ and positive on $(\zeta_{r,0},1)$, we have
\begin{align}
\label{eq:Ar_bound}
A_r(z_0) \le \int_{\zeta_{r,0}}^1h_r(z)\dd z =E_{r,0} \le
E_{n-1,m(\theta)},
\end{align}
where the last inequality follows from Lemmas \ref{lem:r_dominance} and \ref{lem:t_dominance}.

Consider first an extreme point of $\mathcal A(z_0)$ with one positive component. The component cannot be $\alpha_n$, because $h_n(z_0)=\frac{1}{n}>0$ violates the entry constraint. Therefore, it is a component with $r\in\{1,\ldots,n-1\}$, and (\ref{eq:Ar_bound}) gives the result.

Next, suppose that two components $r,s\in\{1,\ldots,n-1\}$ are positive. Then the objective is a convex combination of $A_r(z_0)$ and $A_s(z_0)$, each bounded by $E_{n-1,m(\theta)}$ due to (\ref{eq:Ar_bound}).

Finally, suppose the positive components are $\alpha_n$ and $\alpha_r$, with $r\in\{1,\ldots,n-1\}$. Because the entry constraint binds, the corresponding prize schedule can be written as ${\bf v}^{r,t}$ for some $t\ge 0$ satisfying
\[
\psi(\Phi_r(z_0))+t=0.
\]
Therefore, it must be that $t\le m(\theta)$. If $z_0$ is the larger root of $H_{r,t}(z)$, then
\[
\int_{z_0}^1H_{r,t}(z)\dd z=E_{r,t}.
\]
If $z_0$ is the smaller root,\footnote{The two-prize schedule considered here is an extreme point of $\mathcal A(z_0)$, but it does not necessarily generate a regular equilibrium with cutoff $z_0$. Therefore, because the entry constraint binds, $z_0$ is a root of $H_{r,t}(z)$, but it is not necessarily its largest root.} $H_{r,t}(z)$ is negative between the two roots, and hence
\[
\int_{z_0}^1H_{r,t}(z)\dd z<E_{r,t}.
\]
In either case, 
\[
\int_{z_0}^1H_{r,t}(z)\dd z \le E_{r,t} \le E_{n-1,m(\theta)}.
\]
due to Lemmas \ref{lem:r_dominance} and \ref{lem:t_dominance}, as required.
\end{proof}

\bigskip

\begin{proof}{\bf of Proposition \ref{prop:universal_two_prize_linear}}
Fix any partial-entry CPE, and let $z_0$ be the mass of nonparticipants after any zero-expenditure entrants have been reclassified as nonparticipants. By Lemma \ref{lem:pooling_total},
\[
\mathds{E}[X({\bf v})] \le \int_{z_0}^1H(z;{\bf v})\dd z.
\]
The corresponding budget shares ${\boldsymbol\alpha}$ belong to $\mathcal A(z_0)$ because $H(z_0;{\bf v})\le 0$. For a fixed $z_0$, the right-hand side is linear in ${\boldsymbol\alpha}$ and is, therefore, maximized at an extreme point of $\mathcal A(z_0)$. Lemma \ref{lem:extreme_point_dom} bounds expenditure at any such extreme point by $E_{n-1,m(\theta)}$, and the regular equilibrium under ${\bf v}^{n-1,m(\theta)}$ attains this bound.
\end{proof}

\bigskip

\begin{proof}{\bf of Lemma \ref{lem:positive_rent_dominated}}
Let $u>0$ be the equilibrium utility, and define
\[
s=\min\{u,v_n\}, \qquad R=1-ns.
\]
If $R=0$, then $s=v_n=\frac{1}{n}$, so all prizes are equal and equilibrium expenditure is zero. We, therefore, assume $R>0$. Define rescaled prizes
\begin{align}
\label{eq:rent_removed_prizes}
\tilde v_r=\frac{v_r-s}{R}, \qquad r=1,\ldots,n.
\end{align}
Since $s\le v_n$, $\tilde{\bf v}$ is nonnegative, decreasing, and satisfies the budget constraint $\sum_{r=1}^n\tilde v_r=1$.

Let $B_{\bf v}(x;F)$ denote the benefit for an agent choosing expenditure $x$ under prize schedule ${\bf v}$, assuming that all other agents randomize expenditure according to $F$.

For every original equilibrium expenditure $x$, consider the transformed expenditure
\begin{align}
\label{eq:effort_transform}
\tilde x=T(x)=\frac{x+u-s}{R}.
\end{align}
Transformation $T$ is strictly increasing and, therefore, preserves all rankings and ties.

We argue that if $F$ is an equilibrium expenditure distribution under ${\bf v}$, with corresponding random expenditure $X$, then $\tilde F$---the distribution of $\tilde X=T(X)$---is an equilibrium expenditure distribution under $\tilde{\bf v}$. In particular, expenditure $T(x)$ against $\tilde F$ generates exactly the same rank lottery as expenditure $x$ against $F$. Subtracting $s$ from every prize reduces expected monetary benefit by $s$ and leaves all prize differences, and hence the gain-loss utility, unchanged; division by $R$ scales both components by $\frac{1}{R}$. Hence,
\[
B_{\tilde{\bf v}}(T(x);\tilde F) = \frac{B_{\bf v}(x;F)-s}{R}.
\]

For every $x$ in the support of the original equilibrium, $B_{\bf v}(x;F)-x=u$; therefore,
\begin{align*}
B_{\tilde{\bf v}}(T(x);\tilde F)-T(x) = \frac{B_{\bf v}(x;F)-s}{R} - \frac{x+u-s}{R} = \frac{B_{\bf v}(x;F)-x-u}{R} =0.
\end{align*}
Thus, all expenditures in the support of $\tilde F$ yield zero utility.

We will now consider deviations in the transformed game. Suppose first that $y\ge\frac{u-s}{R}$, and let $x=T^{-1}(y)=Ry-(u-s)\ge 0$. Expenditure $y$ against $\tilde F$ generates the same rank lottery as expenditure $x$ against $F$. Therefore,
\[
B_{\tilde{\bf v}}(y;\tilde F)-y = \frac{B_{\bf v}(x;F)-x-u}{R} \le 0,
\]
where the inequality follows from optimality of the original equilibrium strategy.

Now consider deviations $0\le y<\frac{u-s}{R}$, which are only possible if $u>s$, which implies $s=v_n<u$. Such an expenditure lies strictly below every transformed equilibrium expenditure; therefore, it is ranked last with certainty and generates utility
\[
\tilde v_n-y = \frac{v_n-s}{R}-y = -y \le 0.
\]
Thus, $\tilde F$ is a zero-rent full-entry symmetric mixed CPE under $\tilde{\bf v}$.

Finally, 
\[
T(x)-x = \frac{x+u-s}{R} - x = \frac{u-s+x(1-R)}{R} \ge 0.
\]
Hence, cost expenditure in the transformed equilibrium is higher state by state.
\end{proof}

\bigskip

\begin{proof}{\bf of Corollary \ref{cor:unrestricted_optimum_strong}}
By Lemma \ref{lem:positive_rent_dominated}, it is sufficient to consider zero-rent full-entry CPEs.

First suppose there is a positive mass of agents choosing zero expenditure, with quantile block $[0,z_0]$. Since these agents' expenditure and equilibrium utility are both zero,
\[
\bar H(0,z_0;{\bf v})=0.
\]
An arbitrarily small strictly positive expenditure beats the entire zero-expenditure atom, so equilibrium requires $H(z_0;{\bf v})\le 0$. Reclassify these zero-expenditure agents as nonparticipants. All rank lotteries and payoffs with strictly positive expenditures are unchanged; a deviation to zero now gives $H(z_0;{\bf v})\le 0$, and expected expenditure is unchanged. The resulting strategy profile is a partial-entry CPE, so Proposition \ref{prop:universal_two_prize_linear} applies.

Now suppose there is no atom at zero. We first show that $v_n=0$. If the minimum equilibrium expenditure is strictly positive, deviation to zero expenditure gives utility $v_n$, so zero rent requires $v_n=0$. If the infimum of equilibrium support is zero, take a sequence of supported expenditures converging to zero. Since there is no atom at zero, the associated rank lottery converges to losing with certainty, so equilibrium indifference gives $v_n=0$ again.

With $v_n=0$, let $\alpha_r=rd_r$, so $\alpha_r\ge0$ and $\sum_{r=1}^{n-1}\alpha_r=1$. The same argument as in Lemma \ref{lem:pooling_total}, now over $[0,1]$, gives
\[
\mathds{E}[X({\bf v})]
\le
\int_0^1H(z;{\bf v})\dd z
=
\sum_{r=1}^{n-1}\alpha_r A_r(0).
\]
By (\ref{eq:Ar_bound}), $A_r(0)\le E_{n-1,m(\theta)}$ for every $r$, and hence
\[
\mathds{E}[X({\bf v})]
\le
E_{n-1,m(\theta)}.
\]
The regular partial-entry equilibrium under ${\bf v}^{n-1,m(\theta)}$ attains the bound.
\end{proof}

\bigskip

\begin{proof}{\bf of the expression for $E^*(\theta)$ used in Corollary \ref{cor:maximal_effort}}
For $\theta\in[0,1]$, we have
\[
E^*(\theta) = \frac1{n-1}\int_0^1\psi(\Phi_{n-1}(z))\dd z,
\]
where $\Phi_{n-1}(z)=1-(1-z)^{n-1}$ and hence
\[
\int_0^1\Phi_{n-1}(z)\dd z=\frac{n-1}{n}
\]
and
\[
\int_0^1\Phi_{n-1}(z)^2\dd z
=1-\frac2n+\frac1{2n-1}
=\frac{2(n-1)^2}{n(2n-1)},
\]
which gives the first expression in (\ref{eq:E_star}).

For $\theta>1$, let $\mu^*$ be given by (\ref{eq:optimal_entry_main}). Using $u=1-z$ and
$(\mu^*)^{n-1}=\frac{\theta+1}{2\theta}$, we obtain
\begin{align*}
E_{n-1,m(\theta)} = \frac{\theta}{n-1+nm(\theta)}\int_0^{\mu^*}\left(\frac{\theta+1}{2\theta}-u^{n-1}\right)^2\dd u = \frac{\theta(\mu^*)^{2n-1}}{n-1+nm(\theta)}
\left(1-\frac2n+\frac1{2n-1}\right),
\end{align*}
which gives the second expression in (\ref{eq:E_star}).
\end{proof}

\bigskip

\begin{proof}{\bf of Lemma \ref{lem:optimal_effort_theta}}
For $\theta\in[0,1]$, maximal expected effort is
\[
E^*(\theta) = \frac{2n-1-\theta}{n(2n-1)}.
\]
Therefore,
\[
\frac{\mathrm{d}E^*(\theta)}{\mathrm{d}\theta} = -\frac{1}{n(2n-1)}
<0.
\]
For $\theta>1$, we have
\[
E^*(\theta) = \frac{2\theta(n-1)^2}{n(2n-1)[n-1+n m(\theta)]}\left(\frac{\theta+1}{2\theta}\right)^{\frac{2n-1}{n-1}},
\]
where
\[
m(\theta) = \frac{(\theta-1)^2}{4\theta},
\]
and hence
\[
n-1+n m(\theta) = \frac{n\theta^2+2(n-2)\theta+n}{4\theta}.
\]
Logarithmic differentiation gives
\begin{equation}
\label{eq:dlog_Estar}
\frac{\mathrm{d}\log E^*(\theta)}{\mathrm{d}\theta}=
-\frac{(5n-4)\theta^2-2n\theta+n
}{(n-1)\theta(\theta+1)
\left[n\theta^2+2(n-2)\theta+n\right]
}.
\end{equation}
The denominator in (\ref{eq:dlog_Estar}) is strictly positive. The quadratic polynomial in the numerator, has a strictly positive coefficient on $\theta^2$ and discriminant
\[
(-2n)^2-4(5n-4)n = -16n(n-1)<0.
\]
The numerator is, therefore, strictly positive for all $\theta$. Thus, $\frac{\mathrm{d}E^*(\theta)}{\mathrm{d}\theta}<0$ for $\theta>1$. At $\theta=1$, the two expressions for maximal expected effort coincide, i.e., $E^*(\theta)$ is continuous at $\theta=1$ and, therefore, continuous and strictly decreasing for all $\theta\ge0$.

Finally, for $\theta\to\infty$ we have $m(\theta)\sim\frac{\theta}{4}$ and $
\frac{\theta+1}{2\theta}\to\frac{1}{2}$. It follows that
\begin{align*}
\lim_{\theta\to\infty}E^*(\theta) = \frac{8(n-1)^2}{n^2(2n-1)}
\left(\frac{1}{2}\right)^{\frac{2n-1}{n-1}} = \frac{2^{\frac{n-2}{n-1}}(n-1)^2}{n^2(2n-1)} > 0,
\end{align*}
which proves (\ref{eq:optimal_effort_limit}).
\end{proof}

\bigskip

\begin{proof}{\bf of Lemma \ref{lem:worst_inverse_cost}}
Because the infimum over $\mathcal Q_A$ equals the infimum over its closure, it is enough to minimize over weakly increasing concave functions. Any such function $q:[0,1]\to\mathds{R}_+$ with $q(0)=0$ admits the representation
\[
q(x)=\int_{(0,1]}\min\{x,t\}\,\nu(\dd t)
\]
for a nonnegative measure $\nu$. Conversely, every function of this form is weakly increasing and concave. Tonelli's theorem allows us to swap the order of integration to obtain
\[
\int_0^1q(x)\dd x = \int_{(0,1]}\left[\int_0^1\min\{x,t\}\dd x\right]\nu(\dd t) = \int_{(0,1]}\left(t-\frac{t^2}{2}\right)\nu(\dd t)
\]
and
\[
\mathds{E}[q(X)] = \int_{(0,1]}\mathds{E}[\min\{X,t\}]\nu(\dd t).
\]
Because both expressions are homogeneous in $\nu$, the productivity constraint in (\ref{eq:Q_A}) binds at the infimum. Therefore, the ratio of the two integrals is bounded below by the infimum of the pointwise ratio. A measure concentrated at a minimizing $t$, when one exists, attains this bound in the closure of $\mathcal Q_A$; otherwise, a sequence of measures concentrated at a minimizing sequence of truncations approaches it. Since $\mathcal Q_A$ is dense in its closure, this gives (\ref{eq:worst_inverse_cost}).

For the pointwise inequality, apply (\ref{eq:worst_inverse_cost}) to the degenerate random variable $X=x\le \frac{1}{2}$:
\[
\inf_{q\in\mathcal Q_A}q(x) = A\inf_{t\in(0,1]}\frac{\min\{x,t\}}{t-\frac{t^2}{2}}.
\]
For $t\ge x$, the ratio is minimized at $t=1$ and equals $2x$. For $t<x$, the ratio equals $\frac{1}{1-\frac{t}{2}}\ge 1\ge 2x$. Hence the infimum is $2Ax$, proving (\ref{eq:pointwise_linear_bound}).
\end{proof}

\bigskip

\begin{proof}{\bf of Lemma \ref{lem:linear_least_favorable}}
For $n\ge3$, the claim follows from (\ref{eq:pointwise_linear_bound}) and the bound 
\[
X^*(\theta)\le v_1^*(\theta)\le\frac{1}{n-1}\le\frac{1}{2},
\]
which shows that all realizations of $X^*(\theta)$ lie in the interval where (\ref{eq:pointwise_linear_bound}) applies. Thus, $q(X^*(\theta))\ge2AX^*(\theta)$ state by state for every $q\in\mathcal Q_A$, while equality is attained by $q_L(x)=2Ax$.

It remains to consider $n=2$, for which it is sufficient to show that
\begin{align}
\label{eq:n2_stoploss}
\mathds{E}[\min\{X^*(\theta),t\}] \ge \mathds{E}[X^*(\theta)](2t-t^2),
\qquad t\in(0,1].
\end{align}
Indeed, since $t-\frac{t^2}{2}>0$ for $t\in(0,1]$, inequality (\ref{eq:n2_stoploss}) is equivalent to
\[
\frac{\mathds{E}[\min\{X^*(\theta),t\}]}{t-\frac{t^2}{2}}\ge 2\mathds{E}[X^*(\theta)].
\]
Lemma \ref{lem:worst_inverse_cost} then implies
\[
\inf_{q\in\mathcal Q_A}\mathds{E}[q(X^*(\theta))] \ge 2A\mathds{E}[X^*(\theta)],
\]
and since $q_L(x)=2Ax\in\mathcal Q_A$, the reverse inequality holds as well.

Let $x^*(z;\theta)$ denote the equilibrium expenditure quantile function, so that
\[
X^*(\theta)=x^*(Z;\theta), \qquad Z\sim{\rm Uniform}[0,1].
\]
For $\theta\in[0,1]$, ${\bf v}^*=(1,0)$ and $x^*(z;\theta) = \psi(z) = (1-\theta)z+\theta z^2$. This gives
\[
\mathds{E}[X^*(\theta)] = \int_0^1x^*(z;\theta)\dd z = \frac{1-\theta}{2} + \frac{\theta}{3} = \frac{3-\theta}{6}.
\]
Let $t=x^*(s;\theta)$, $s\in[0,1]$. Since $x^*(z;\theta)$ is increasing in $z$,
\[
\mathds{E}[\min\{X^*(\theta),t\}] = \int_0^sx^*(z;\theta)\dd z+t(1-s) = \frac{(1-\theta)s^2}{2} + \frac{\theta s^3}{3} + t(1-s).
\]
This gives
\begin{align*}
& \mathds{E}[\min\{X^*(\theta),t\}] - \mathds{E}[X^*(\theta)](2t-t^2) = \frac{(1-\theta)s^2}{2} + \frac{\theta s^3}{3} + t(1-s) - \frac{(3-\theta)(2t-t^2)}{6}\\
& = \frac{(1-\theta)s^2}{2} + \frac{\theta s^3}{3} + [(1-\theta)s+\theta s^2](1-s) - \frac{3-\theta}{6}[(1-\theta)s+\theta s^2][2-(1-\theta)s-\theta s^2]\\
& = \frac{s}{6}[3(1-\theta)s+2\theta s^2+(6-6\theta+6\theta s)(1-s)-(3-\theta)(1-\theta+\theta s)(2-(1-\theta)s-\theta s^2)]
\end{align*}
The last expression in brackets, which we denote by $B$, can be further simplified by introducing $r=1-s$:
\begin{align*}
& B = 3(1-\theta)(1-r)+2\theta(1-r)^2+6r(1-\theta r) - (3-\theta)(1-\theta r)[1+(1+\theta)r-\theta r^2] \\
& = 3(1-\theta) - 3r(1-\theta) + 2\theta - 4\theta r + 2\theta r^2 + 6r - 6\theta r^2 - [3-\theta-\theta(3-\theta)r][1+(1+\theta)r-\theta r^2] \\
& = 3-\theta + (3-\theta)r - 4\theta r^2 - [3 - \theta - \theta(3-\theta)r +(3-\theta)(1+\theta)r -\theta(3-\theta)(1+\theta)r^2 \\
& - \theta(3-\theta)r^2 + \theta^2(3-\theta)r^3] = - 4\theta r^2 + \theta(3-\theta)(1+\theta)r^2 + \theta(3-\theta)r^2 - \theta^2(3-\theta)r^3 \\
& = \theta r^2[-4 + (3-\theta)(1+\theta) + 3- \theta -\theta(3-\theta)r] = \theta r^2[2 + \theta - \theta^2 - \theta(3-\theta)r].
\end{align*}
Plugging in $r=1-s$, we finally have
\[
B = \theta(1-s)^2[2 + \theta - \theta^2 - \theta(3-\theta)(1-s)] = \theta(1-s)^2[2(1-\theta)+s\theta(3-\theta)]\ge 0,
\]
where the inequality follows from $\theta\in[0,1]$ and $s\in[0,1]$. The expression is identically zero at $\theta=0$. This proves (\ref{eq:n2_stoploss}) for moderate loss aversion.

We now consider $\theta>1$. For $n=2$, we have $\Phi_{n-1}(z)=\Phi_1(z)=z$; therefore, under the optimal prize schedule the entry cutoff is
\[
z_0=P_m(\theta)=\frac{\theta-1}{2\theta},
\]
and hence the equilibrium entry probability is
\[
\mu=1-z_0=\frac{\theta+1}{2\theta}.
\]
Using (\ref{eq:H_optimal_main}), we have under the optimal prize schedule
\[
H(z;{\bf v}^*(\theta)) = \frac{\theta(z-z_0)^2}{1+2m(\theta)}.
\]
Since
\[
1+2m(\theta) = 1+\frac{(\theta-1)^2}{2\theta} = \frac{\theta^2+1}{2\theta},
\]
we obtain
\[
H(z;{\bf v}^*(\theta)) = K(z-z_0)^2, \qquad K = \frac{2\theta^2}{\theta^2+1}.
\]

The optimal equilibrium is a regular partial-entry equilibrium with zero rent. Thus, agents with effective quantiles below $z_0$ do not participate and have zero expenditure, while for $z\ge z_0$ equilibrium expenditure equals $H(z;{\bf v}^*(\theta))$. Hence,
\[
x^*(z;\theta)
=
\begin{cases}
0, & z<z_0,\\
\displaystyle
K(z-z_0)^2,
& z\ge z_0.
\end{cases}
\]
Expected equilibrium expenditure is
\[
\mathds{E}[X^*(\theta)] = \int_{z_0}^1K(z-z_0)^2\dd z = \frac{K\mu^3}{3}.
\]
Now consider any $t$ in the support of $X^*(\theta)$. Since $x^*(z;\theta)$ is strictly increasing for $z\ge z_0$, there exists $s\in[z_0,1]$ such that
\[
t=x^*(s;\theta)=K(s-z_0)^2.
\]
For $w\in[0,1]$, define $s=z_0+w\mu$. Then $s-z_0=w\mu$, $1-s=\mu(1-w)$, and hence $t=Kw^2\mu^2$. Therefore,
\begin{align*}
& \mathds{E}[\min\{X^*(\theta),t\}] = \int_{z_0}^sK(z-z_0)^2\dd z+t(1-s) =
\frac{K(s-z_0)^3}{3} + Kw^2\mu^2(1-s) \\
& = \frac{Kw^3\mu^3}{3} + Kw^2\mu^3(1-w) = K\mu^3\left(w^2-\frac{2w^3}{3}\right).
\end{align*}
This gives
\begin{align*}
& \mathds{E}[\min\{X^*(\theta),t\}] - \mathds{E}[X^*(\theta)](2t-t^2) = K\mu^3\left(w^2-\frac{2w^3}{3}\right) - \frac{K\mu^3}{3}\left(2Kw^2\mu^2-K^2w^4\mu^4\right) \\
& = \frac{K\mu^3w^2}{3}\left[3-2w-2K\mu^2+K^2\mu^4w^2\right].
\end{align*}
Substituting the expressions for $K$ and $\mu$, we have
\begin{align*}
& 3-2w-2K\mu^2+K^2\mu^4w^2 = \frac{1}{4(\theta^2+1)^2}\big[(\theta+1)^4w^2 - 8(\theta^2+1)^2w 
+ 12(\theta^2+1)^2 \\
& - 4(\theta+1)^2(\theta^2+1)\big] = \frac{1}{4(\theta^2+1)^2}P_\theta(w),
\end{align*}
where
\[
P_\theta(w) = (\theta+1)^4w^2 - 8(\theta^2+1)^2w + 8(\theta^4-\theta^3+2\theta^2-\theta+1).
\]
Thus,
\begin{align*}
&\mathds{E}[\min\{X^*(\theta),t\}] - \mathds{E}[X^*(\theta)](2t-t^2) = \frac{w^2(\theta+1)^3}
{48\theta(\theta^2+1)^3}P_\theta(w).
\end{align*}

For $\theta>1$, $P_\theta(w)$ is strictly decreasing on $[0,1]$, because its minimizer
\[
w^* = \frac{4(\theta^2+1)^2}{(\theta+1)^4} = \left[\frac{2(\theta^2+1)}{(\theta+1)^2}\right]^2 > 1.
\]
To prove the last inequality, note that
\[
2(\theta^2+1) - (\theta+1)^2 = (\theta-1)^2 > 0.
\]
Moreover, 
\begin{align*}
& P_\theta(1) = (\theta+1)^4 - 8(\theta^2+1)^2 + 8(\theta^4-\theta^3+2\theta^2-\theta+1) \\
& = \theta^4 + 4\theta^3 + 6\theta^2 + 4\theta + 1 - 8\theta^4 - 16\theta^2 - 8 + 8\theta^4 - 8\theta^3 + 16\theta^2 - 8\theta + 8 \\
& = \theta^4 - 4\theta^3 + 6\theta^2 - 4\theta + 1 = (\theta-1)^4\ge 0; 
\end{align*}
therefore, $P_\theta(w)\ge 0$ and (\ref{eq:n2_stoploss}) holds for all $t$ in the support of $X^*(\theta)$. For $t$ above the maximal realization of $X^*(\theta)$, the numerator in (\ref{eq:worst_inverse_cost}) is constant at $\mathds{E}[X^*(\theta)]$, while $t-\frac{t^2}{2}\le \frac{1}{2}$, so the same inequality follows. Thus, the infimum in (\ref{eq:worst_inverse_cost}) is attained at $t=1$ and equals $2A\mathds{E}[X^*(\theta)]$. The minimizing inverse cost is $q_L(x)=2Ax$.
\end{proof}

\section{Moderate loss aversion with fixed convex costs}
\label{app_additional}

\begin{proposition}
\label{prop_moderate_convex}
Suppose $c$ is convex and $\theta\in[0,1]$. Then the prize schedule ${\bf v}^{n-1}=(\frac{1}{n-1},\ldots,\frac{1}{n-1},0)$ maximizes equilibrium expected effort.
\end{proposition}

\begin{proof}{\bf of Proposition \ref{prop_moderate_convex}}
Recall that $\psi(P)=(1-\theta)P+\theta P^2$. For $r=1,\ldots,n-1$, define
\[
C_r(q)=\frac{1}{r}\psi(\Phi_r(q)),
\]
which is the equilibrium cost quantile under the top-$r$ prize schedule $(\frac{1}{r},\ldots,\frac{1}{r},0,\ldots,0)$.

It is clear from Proposition \ref{prop_eq_moderate} that $v_n=0$ is optimal in this case. For ${\bf v}\in\mathcal{V}$ with $v_n=0$, let $\alpha_r=rd_r\ge 0$. Then the budget constraint becomes $\sum_{r=1}^{n-1}\alpha_r=1$ and
\begin{align}
\label{H_mixture_moderate}
H(q;{\bf v}) = \sum_{r=1}^{n-1}\alpha_r C_r(q).
\end{align}
Since $\theta\le 1$, $\psi(P)$ is increasing on $[0,1]$; therefore, each $C_r(q)$ and $H(q;{\bf v})$ are increasing in $q$.

We first show that, for every $r=1,\ldots,n-2$ and every $p\in[0,1]$,
\begin{align}
\label{Cr_integrated_order}
\int_0^p C_{r+1}(q)\dd q \ge \int_0^p C_r(q)\dd q.
\end{align}
Because of the structure of $\psi$, it is enough to establish the corresponding inequalities separately for $\Phi_r$ and $\Phi_r^2$. Using $\Phi_{r+1}(q)=\Phi_r(q)+\phi_{r+1}(q)$ and (\ref{phi_Phi_identity}), and integrating by parts,
\begin{align*}
\int_0^p\Phi_{r+1}(q)\dd q = \int_0^p\Phi_r(q)\dd q + \frac{1}{r}\int_0^p(1-q)\Phi_r'(q)\dd q = \frac{r+1}{r}\int_0^p\Phi_r(q)\dd q + \frac{1-p}{r}\Phi_r(p).
\end{align*}
This gives
\begin{align}
\label{Phi_integrated_order}
\frac{1}{r+1}\int_0^p\Phi_{r+1}(q)\dd q - \frac1r\int_0^p\Phi_r(q)\dd q =
\frac{(1-p)\Phi_r(p)}{r(r+1)} \ge 0.
\end{align}
Similarly,
\begin{align*}
& \int_0^p\Phi_{r+1}(q)^2\dd q = \int_0^p \left[\Phi_r(q)+\phi_{r+1}(q)\right]^2\dd q\\
& = \int_0^p\Phi_r(q)^2\dd q + 2\int_0^p\Phi_r(q)\phi_{r+1}(q)\dd q + \int_0^p\phi_{r+1}(q)^2\dd q.
\end{align*}
Using (\ref{phi_Phi_identity}),
\begin{align*}
2\int_0^p\Phi_r(q)\phi_{r+1}(q)\dd q = \frac{1}{r}\int_0^p (1-q)[\Phi_r(q)^2]'\dd q
= \frac{1}{r}\left[(1-p)\Phi_r(p)^2 + \int_0^p\Phi_r(q)^2\dd q \right].
\end{align*}
It follows that
\begin{align}
\label{Phi2_integrated_order}
\frac{1}{r+1}\int_0^p\Phi_{r+1}(q)^2\dd q - \frac1r\int_0^p\Phi_r(q)^2\dd q = \frac{(1-p)\Phi_r(p)^2}{r(r+1)} + \frac{1}{r+1}\int_0^p\phi_{r+1}(q)^2\dd q \ge 0.
\end{align}
Multiplying (\ref{Phi_integrated_order}) by $1-\theta$, multiplying
(\ref{Phi2_integrated_order}) by $\theta$, and adding establishes
(\ref{Cr_integrated_order}). Iterating (\ref{Cr_integrated_order}) gives, for every
$r=1,\ldots,n-2$ and $p\in[0,1]$,
\[
\int_0^p C_{n-1}(q)\dd q \ge \int_0^p C_r(q)\dd q.
\]
Using (\ref{H_mixture_moderate}) and $\sum_{r=1}^{n-1}\alpha_r=1$, we, therefore, obtain
\begin{align}
\label{H_SSD}
\int_0^p C_{n-1}(q)\dd q \ge \sum_{r=1}^{n-1}\alpha_r \int_0^p C_r(q)\dd q =
\int_0^p H(q;{\bf v})\dd q \quad
\text{for all }p\in[0,1].
\end{align}
Since $c$ is increasing and convex, $c^{-1}$ is increasing and concave. For $Z\sim{\rm Uniform}[0,1]$, condition (\ref{H_SSD}) shows that $C_{n-1}(Z)$ dominates $H(Z;{\bf v})$ in the increasing concave order. Therefore,
\[
\int_0^1c^{-1}(C_{n-1}(q))\dd q \ge \int_0^1c^{-1}(H(q;{\bf v}))\dd q.
\]
By Proposition \ref{prop_eq_moderate}, the left-hand side is expected effort under ${\bf v}^{n-1}$ and the right-hand side is expected effort under ${\bf v}$. Therefore, ${\bf v}^{n-1}$ is optimal.
\end{proof}

\end{document}